\documentclass[british]{scrartcl}
\usepackage[T1]{fontenc}
\usepackage[latin9]{inputenc}
\usepackage[a4paper]{geometry}
\usepackage{babel}
\usepackage{verbatim}
\usepackage{amsmath}
\usepackage{graphicx}
\usepackage[unicode=true,
 bookmarks=true,bookmarksnumbered=false,bookmarksopen=false,
 breaklinks=true,pdfborder={0 0 0},backref=false,colorlinks=false]
 {hyperref}
\hypersetup{pdftitle={Pollinator foraging adaptation and the coexistence of competing plants},
 pdfauthor={Tomas Revilla},
 pdfkeywords={mutualism, apparent competition, optimal foraging, coexistence, adaptation delay}}
\usepackage{breakurl}

\makeatletter

\providecommand{\tabularnewline}{\\}

\usepackage{lineno}
\usepackage[labelfont=bf,font=footnotesize,labelsep=period,format=plain]{caption}
\usepackage{setspace}

\usepackage[auth-lg]{authblk}
\author[1]{Tom\'{a}s A. Revilla\footnote{corresponding author, email: tomrevilla@gmail.com}}
\author[1,2]{Vlastimil K\v{r}ivan\footnote{email: vlastimil.krivan@gmail.com}}
\affil[1]{Institute of Entomology, Biology Center, Czech Academy of Sciences, 
                Brani\v{s}ovsk\'{a} 31, 370 05 \v{C}esk\'{e} Bud\v{e}jovice, Czech Republic}
\affil[2]{Department of Mathematics and Biomathematics, Faculty of Science, University of South Bohemia, 
                Brani\v{s}ovsk\'{a} 31, 370 05 \v{C}esk\'{e} Bud\v{e}jovice, Czech Republic}

\makeatother

\begin{document}

\title{Pollinator foraging adaptation and coexistence of competing plants}
\maketitle
\begin{abstract}
We use the optimal foraging theory to study coexistence between two plant species
and a generalist pollinator. We compare conditions for plant coexistence for non-adaptive
vs. adaptive pollinators that adjust their foraging strategy to maximize fitness.
When pollinators have fixed preferences, we show that plant coexistence typically
requires both weak competition between plants for resources (e.g., space or nutrients)
and pollinator preferences that are not too biased in favour of either plant. We
also show how plant coexistence is promoted by indirect facilitation via the pollinator.
When pollinators are adaptive foragers, pollinator's diet maximizes pollinator's
fitness measured as the per capita population growth rate. Simulations show that
this has two conflicting consequences for plant coexistence. On the one hand, when
competition between pollinators is weak, adaptation favours pollinator specialization
on the more profitable plant which increases asymmetries in plant competition and
makes their coexistence less likely. On the other hand, when competition between
pollinators is strong, adaptation promotes generalism, which facilitates plant coexistence.
In addition, adaptive foraging allows pollinators to survive sudden loss of the preferred
plant host, thus preventing further collapse of the entire community.

\end{abstract}

\section{Introduction}
\begin{quote}
\begin{flushleft}
\emph{Et il se sentit très malheureux. Sa fleur lui avait raconté qu'elle était seule
de son espèce dans l'univers. Et voici qu'il en était cinq mille, toutes semblables,
dans un seul jardin!}
\par\end{flushleft}

\begin{flushleft}
Le Petit Prince, Chapitre XX -- Antoine de Saint-Exupéry
\par\end{flushleft}
\end{quote}
The diversity and complexity of mutualistic networks motivate ecologists to investigate
how they can remain stable and persistent over time. Mathematical models and simulations
show that some properties of mutualistic networks (e.g., low connectance and high
nestedness) make them more resistant against cascading extinctions \cite{memmott_etal-rspb04},
more likely to sustain large numbers of species \cite{bastolla_etal-nature09}, and
more stable demographically \cite{okuyama_holland-ecolett08}. However, simulations
\cite{benadi_etal-amnat13,bewick_etal-oikos13} also indicate that mutualism increases
competitive asymmetries, causing complex communities to be less persistent. These
studies consider large numbers of species, parameters and initial conditions, making
it difficult to understand the interplay between mutualisms (e.g., between plant
and animal guilds) and antagonisms (e.g., resource competition between plants). These
questions are easier to study in the case of community modules consisting of a few
species only \cite{holt-97}.

\begin{figure}[t]
\begin{centering}
\includegraphics[clip]{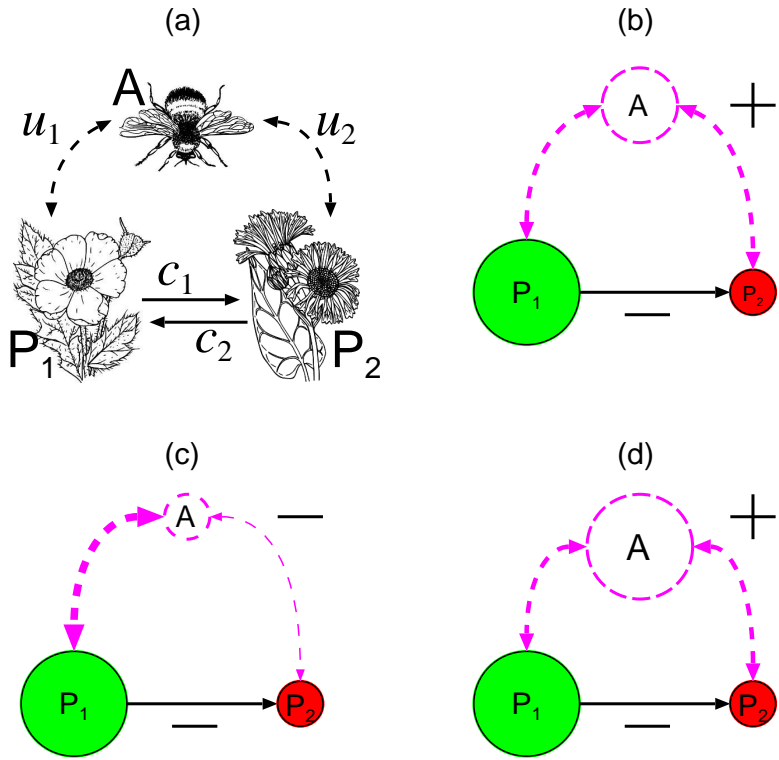} 
\par\end{centering}

\protect\caption{\label{fig:model}\textbf{Community module consisting of plant 1 and 2, and pollinator
A.} (a) Plants affect each other directly (solid arrows) by competition for space
or resources $(c_{1},c_{2})$, and indirectly (dashed arrows) via shared pollinator
with plant preferences $u_{1}$ and $u_{2}$. (b) When pollinator preferences are
fixed and not too biased, a large density of plant 1 maintains a large pollinator
density, which has an indirect positive effect on low density plant 2. In (c,d) pollinator
preferences for plants are adaptive (dashed arrows change thickness). When pollinators
are rare (c), preferences favour abundant plant 1, which results in a negative indirect
effect on rare plant 2. When pollinators become abundant (d), competition between
pollinators lead to balanced preferences, and the indirect effect on plant 2 becomes
positive. The viability of plant 2 depends on the balance between indirect and direct
effects. Image sources for panel (a) were taken from: https://openclipart.org.}
\end{figure}

In this article we consider a mutualistic module with two plant species and one pollinator
species (Fig \ref{fig:model}a). This module combines several direct and indirect
interactions that are either density- or trait-mediated (sensu \cite{bolker_etal-ecology03}).
These include plant intra- and inter-specific competition (for e.g., space), plant
competition for pollinator services, and pollinator intra-specific competition for
plant resources (e.g., nectar). Some of these interactions depend on changes in population
densities only (e.g., intra- and inter-specific plant competition), while the others
depend also on individual morphological and behavioural traits. As some of them have
positive and some of them negative effect on plant coexistence, it is difficult to
predict their combined effects on species persistence and stability.

First, we will assume that pollinator preferences for plants are fixed. In this case,
there is a negative effect of one plant on the other by direct competition and a
positive indirect effect that is mediated by the shared pollinators, called facilitation
\cite{rathcke-real1983,moeller-ecology04,ghazoul-joe06}. As one plant population
density increases, pollinator density increases too, which, in turn, increases pollination
rate of the other plant (Fig \ref{fig:model}b). This is an indirect interaction
between plants that is mediated by changes in abundance of the pollinator (i.e.,
density mediated indirect interaction). Because facilitation has the opposite effect
to direct plant competition (see Fig \ref{fig:model}b) it is important to clarify
under which situations the positive effect of facilitation prevails, and we study
this question by using a mathematical model.

Second, we will assume that pollinator preferences are adaptive. We will assume that
pollinator fitness is defined as the per capita pollinator population growth rate
that depends on plant (that produce resources for pollinators) as well as on pollinator
densities. First, pollinators benefit from nectar quality and nectar abundance (which
correlates with plant population density). Second, pollinators compete for resources.
This competition will play an important effect when pollinator population densities
are high. A game theoretical approach to determine the optimal pollinator strategy
is the Ideal Free Distribution (IFD) \cite{fretwell_lucas-actabiot69,krivan_etal-tpb08}.
This theory predicts that when pollinators are at low numbers, they will specialize
on one plant only. As their population density will increase, they become generalists
feeding on and pollinating both plants. This mechanism causes a negative effect of
the preferred plant on the other plant, because when at low densities, pollinators
will specialize on one plant only (Fig \ref{fig:model}c). This is an example of
a positive feedback where ``the rich becomes richer and the poor get poorer''.
Competition for pollinators is an example of a trait-mediated effect caused by pollinator
behaviour. Pollinator specialization on one plant only is detrimental for the other
plant. However, as pollinator population density will increase, competition for resources
among pollinators will increase too \cite{fontaine_etal-joe08}, and the IFD predicts
that they become generalists, which promotes plant coexistence (Fig \ref{fig:model}d).
Once again, combination of positive and negative effects between plants creates complicated
feedbacks between population densities and pollinator behaviour that are impossible
to disentangle without an appropriate mathematical model.

Our main goal is to study how pollinator preferences and plant competition affect
plant coexistence. First, we study the dynamics of the plant--pollinator module when
pollinator preferences are fixed. Second, we calculate the pollinator's evolutionarily
stable foraging strategy (ESS) at fixed plant and pollinator population densities,
and we study plant coexistence assuming pollinators instantaneously track their ESS.
This case corresponds to time scale separation where population dynamics operate
on a slow time scale, while pollinator foraging preferences operate on a fast time
scale. Finally, we consider the situation without time scale separation and we model
preference dynamics with the replicator equation. Overall, we show that pollinator
foraging adaptation has complex effects, sometimes equivocal, on plant coexistence.
On the one hand pollinator adaptation increases competitive asymmetries among plants,
promoting competitive exclusion. On the other hand competition for plant resources
among pollinators promotes generalism over specialization, which can prevent the
loss of pollination services for some plants and promote coexistence.

\section{Methods}

\subsection{Mutualistic community model}

Let us consider two plant species $P_{i}$ ($i=1,2$) and one pollinator species
$A$ (Fig \ref{fig:model}a). Plants produce resources $F_{i}$ ($i=1,2$) such as
nectar at a rate $a_{i}$ per plant. Resources not consumed by the pollinator decrease
with rate $w_{i}$ (e.g., nectar can be re-absorbed, decay or evaporate). Resources
are consumed by pollinators at rate $b_{i}$ per resource per pollinator. Pollinator's
relative preferences for either plant are denoted by $u_{i}$ with $u_{1}+u_{2}=1$.
Plant birth rates are proportional to the rate of pollen transfer that is concomitant
with resource consumption. Thus, we assume that plant birth rates are proportional
to pollinator resource consumption rates ($u_{i}b_{i}F_{i}A$) multiplied by conversion
efficiency $r_{i}$. Pollinator birth rates are proportional to resource consumption
with corresponding conversion efficiency $e_{i}$. Plants and pollinators die with
the per capita mortality rate $m_{i}$ $(i=1,2)$ and $d$, respectively.

Assuming that plant resources equilibrate quickly with current plant and pollinator
densities \cite{revilla-jtb15}, i.e., \emph{$dF_{i}/dt=0$}, plants and pollinator
population dynamics are described by the following model (Supporting Information
\ref{sec:appendixA} Appendix)

\begin{subequations}\label{eq:ode_ss} 
\begin{align}
\frac{dP_{1}}{dt} & =\left(\frac{r_{1}a_{1}u_{1}b_{1}A}{w_{1}+u_{1}b_{1}A}\left(1-\frac{P_{1}+c_{2}P_{2}}{K_{1}}\right)-m_{1}\right)P_{1}\label{eq:ode_pl1_ss}\\
\frac{dP_{2}}{dt} & =\left(\frac{r_{2}a_{2}u_{2}b_{2}A}{w_{2}+u_{2}b_{2}A}\left(1-\frac{P_{2}+c_{1}P_{1}}{K_{2}}\right)-m_{2}\right)P_{2}\label{eq:ode_pl2_ss}\\
\frac{dA}{dt} & =\left(\frac{e_{1}a_{1}u_{1}b_{1}P_{1}}{w_{1}+u_{1}b_{1}A}+\frac{e_{2}a_{2}u_{2}b_{2}P_{2}}{w_{2}+u_{2}b_{2}A}-d\right)A,\label{eq:ode_pol_ss}
\end{align}
\end{subequations}

\noindent in which plant growth rates are regulated by competition for non-living
resources (e.g., light, nutrients, space) according to the Lotka--Volterra competition
model, where $c_{j}$ is the negative effect of plant \emph{j} on plant \emph{i}
relative to the effect of plant \emph{i} on itself (i.e., competition coefficient),
and $K_{i}$ stands for the habitat carrying capacity \cite{benadi_etal-amnat13}.
Notice that plant growth rates saturate with pollinator density (e.g., $\frac{r_{1}a_{1}u_{1}b_{1}A}{w_{1}+u_{1}b_{1}A}$)
and pollinator growth rates decrease due to intra-specific competition for plant
resources (e.g., $\frac{e_{1}a_{1}u_{1}b_{1}P_{1}}{w_{1}+u_{1}b_{1}A}$) \cite{schoener-tpb76}.
In this model plants and pollinators are obligate mutualists, i.e., without pollinators
plants go extinct and without plants the pollinator goes extinct. We do not model
facultative mutualism because this introduces additional factors (e.g. alternative
pollinators, vegetative growth), which complicate the analysis of direct and indirect
effects of the three species module.

\subsection{Fixed pollinator preferences}

We start our analyses assuming that pollinator preferences for plants ($u_{1}$ and
$u_{2}=1-u_{2}$) are fixed at particular values ranging from 0 to 1. This means
that for $u_{1}=1$ or 0 pollinators are plant 1 or plant 2 specialists, respectively,
while for $0<u_{1}<1$ they are generalists. Since model (\ref{eq:ode_ss}) is non-linear,
analytical formulas for interior equilibria and corresponding stability conditions
are out of reach. However, it is possible to obtain coexistence conditions by means
of invasibility analysis. First, we obtain conditions for stable coexistence of a
single plant-pollinator subsystem at an equilibrium. Second, we ask under what conditions
the missing plant species can invade when the resident plant--pollinator subsystem
is at the equilibrium. In particular, we are interested in the situation where each
plant species can invade the other one, because this suggests coexistence of both
plants and pollinators. Derivation of invasion conditions are provided in Supporting
Information \ref{sec:appendixA} Appendix.

In general, invasibility does not guarantee coexistence \cite{armstrong_mcgehee-jtb76,armstrong_mcgehee-amnat80}.
Also, a failure to invade when rare does not rule out possibility of invasion success
when the invading species is at large densities. For these reasons we complement
our invasibility analysis by numerical bifurcation analysis using XPPAUT \cite{xpp-book},
and parameter values given in Table \ref{tab:parameters}. While not empirical, the
values fall within ranges typically employed by consumer--resource models (e.g.,
\cite{grover1997}). Plant-specific parameters are equal except for $e_{i}$ and
$u_{i}$ ($i=1,2$). We assume that $e_{1}>e_{2}$, i.e., plant 1 provides pollinators
with higher energy when compared to plant 2.

\begin{table}
\begin{centering}
\begin{tabular}{clc}
Symbol  & Description  & Values/ranges\tabularnewline
\hline 
$r_{i}$  & conversion efficiency of pollination service to plant $i$ seeds  & 0.1\tabularnewline
$e_{i}$  & conversion efficiency of plant $i$ resources to pollinator eggs  & $e_{1}=0.2,e_{2}=0.1$\tabularnewline
$m_{i}$  & plant $i$ per capita mortality rate  & 0.01\tabularnewline
$d$  & pollinator per capita mortality rate  & 0.1\tabularnewline
$c_{i}$  & competitive effect of plant $i$ on plant $j$  & $c_{i}\geq0$\tabularnewline
$K_{i}$  & plant $i$ habitat carrying capacity  & $K_{i}>0$\tabularnewline
$a_{i}$  & plant $i$ per capita resource production rate  & 0.4\tabularnewline
$b_{i}$  & pollinator consumption rate of plant $i$ resources  & 0.1\tabularnewline
$w_{i}$  & plant $i$ resource decay rate  & 0.25\tabularnewline
$u_{i}$  & relative preference for plant $i$, where $u_{1}+u_{2}=1$  & 0 to 1\tabularnewline
$\nu$  & preference adaptation rate  & $\nu\geq0$\tabularnewline
\hline 
\end{tabular}
\par\end{centering}

\protect\caption{\label{tab:parameters}\textbf{Parameters of model (\ref{eq:ode_ss}) and equation
(\ref{eq:replicator})}. Plant 1 resources are more beneficial for the pollinator
than those from plant 2 $(e_{1}>e_{2})$, but all the other plant-specific parameters
have the same values in order to facilitate comparisons.}
\end{table}

\subsection{Adaptive pollinator preferences}

When pollinators behave as adaptive foragers their plant preferences should maximize
their fitness. The pay-off a pollinator gets when pollinating only plant $i$ is
defined as the per capita pollinator growth rate on that plant, i.e.,

\begin{equation}
V_{1}(u_{1})=\frac{e_{1}a_{1}b_{1}P_{1}}{w_{1}+u_{1}b_{1}A},\hspace{1cm}V_{2}(u_{1})=\frac{e_{2}a_{2}b_{2}P_{2}}{w_{2}+(1-u_{1})b_{2}A}.\label{eq:payoffs}
\end{equation}

We observe that these pay-offs depend both on plant and pollinator densities and
on the pollinator distribution $u_{1}$, i.e., they are both density and frequency
dependent. Now let us consider fitness of a generalist mutant pollinator with strategy
$\tilde{u}_{1}$. Its fitness is then defined as the average pay-off, i.e.,

\begin{equation}
W(\tilde{u}_{1},u_{1})=\tilde{u}_{1}V_{1}(u_{1})+(1-\tilde{u}_{1})V_{2}(u_{1})=V_{2}(u_{1})+\left(V_{1}(u_{1})-V_{2}(u_{1})\right)\tilde{u}_{1}.\label{eq:fitness}
\end{equation}

Using this fitness function we will calculate the evolutionarily stable strategy
\cite{maynardsmith_price-nature73,hofbauersigmund1998} of pollinator preferences
at current plant and pollinator densities. When pollinators adjust their preferences
very fast as compared to changes in population densities, we will use the ESS together
with population dynamics (\ref{eq:ode_ss}) to model effects of pollinator plasticity
on population dynamics. This approach corresponds to time scale separation where
population densities (plants and animals) change very slowly compared to pollinator
adaptation. We are also interested in the situation when the two time scales are
not separated, but pollinators foraging preferences still tend to the ESS. In these
cases we use the replicator equation \cite{hofbauersigmund1998} to model dynamics
of pollinator preferences $u_{1}$ for plant 1 ($u_{2}=1-u_{1}$)

\begin{equation}
\frac{du_{1}}{dt}=\nu u_{1}(1-u_{1})\left(\frac{e_{1}a_{1}b_{1}P_{1}}{w_{1}+u_{1}b_{1}A}-\frac{e_{2}a_{2}b_{2}P_{2}}{w_{2}+(1-u_{1})b_{2}A}\right),\label{eq:replicator}
\end{equation}

\noindent where $\nu\geq0$ is the adaptation rate. Equation (\ref{eq:replicator})
assumes that pollinator's preferences evolve towards a higher energy intake and its
equilibrium coincides with the ESS. Thus, if pollinators obtain more energy when
feeding on plant 1, preferences for plant 1 increases. When $\nu\geq1$, adaptation
is as fast as population dynamics or faster. This describes plastic pollinators that
track changing flower densities very quickly (i.e., within an individual life-span).
This is the case when adaptation is a behavioural trait. In fact, for $\nu$ tending
to infinity pollinators adopt the ESS instantaneously. Adaptation can also involve
morphological changes requiring several generations (i.e., evolution). In that case
$\nu<1$, and adaptation lags behind population dynamics (i.e., changes in preferences
require more generations). And the $\nu=0$ case applies to non-adaptive pollinators.
We remark that perfect specialization on plant 1 or plant 2 correspond to the equilibrium
$u_{1}=1$ or $u_{1}=0$, respectively.

Using model (\ref{eq:ode_ss}) and replicator equation (\ref{eq:replicator}), we
simulate the effects of pollinator adaptation and plant direct inter-specific competition
on coexistence. We consider four common inter-specific competition coefficients:
$c_{1}=c_{2}=c=0$, 0.4, 0.8, and 1.2, and four adaptation rates: $\nu=0$, 0.1,
1 and $\nu=\infty$. Level $\nu=0$ extends our analysis for non-adaptive pollinators
(fixed preferences) beyond invasion conditions. Level $\nu=0.1$ implies slow evolutionary
adaptation, like in adaptive dynamics \cite{geritz_etal-evoeco98}. At $\nu=1$ adaptation
is as fast as demography, i.e., pollinators adapt during their lifetime. For $\nu=\infty$
adaptation is infinitely fast when compared to population densities and preferences
are given by the ESS.

Community dynamics and the dynamics of pollinator preferences can be sensitive to
initial conditions. There are four degrees of freedom for the initial conditions
($P_{1},P_{2},A$ and $u_{1}$ at $t=0$). We reduce this number to two degrees of
freedom. First, we vary $P_{1}(0)$ from 0 to $K$ in 100 steps while $P_{2}(0)=K-P_{1}(0)$,
where $K=K_{1}=K_{2}=50$ is the common carrying capacity. The choice $K=50$ is
high enough to avoid pollinator extinction due to the Allee effect in the majority
of the simulations. Second, we consider two scenarios:
\begin{description}
\item [{Scenario~I:}] Initial pollinator density $A(0)$ varies from 0 to 50 in 100 steps
and initial pollinator preference is equal to the ESS.
\item [{Scenario~II:}] Initial pollinator preference $u_{1}(0)$ varies from 0.001 to
0.999 in 100 steps {[}0.001, 0.01, 0.02, \ldots{}, 0.98, 0.99, 0.999{]} and initial
pollinator density is kept at $A(0)=2$.
\end{description}
Scenario \textbf{I} assumes that pollinators preferences are at the ESS for given
initial plant and pollinator densities, with an exception when the ESS is 0 or 1
in which case we perturb it to $u_{1}=0.001$ or $u_{1}=0.999$. This is necessary
because the replicator equation (\ref{eq:replicator}) does not consider mutations
that may allow specialists to evolve towards generalism.

Scenarios \textbf{I} and \textbf{II} complement each other. In both of them initial
plant composition $(P_{1}:P_{2})$ influences the outcome. For scenario \textbf{II}
we also used $A(0)=50$, but we did not find important qualitative differences. Thus,
for both scenarios we simulate model (\ref{eq:ode_ss}) and (\ref{eq:replicator})
with $100\times100=10^{4}$ different initial conditions. This systematic approach
allows us to delineate boundaries between plant coexistence and extinction regions.
Model (\ref{eq:ode_ss}) and (\ref{eq:replicator}) is integrated (Runge--Kutta 4th,
with Matlab \cite{MATLAB2010}) with the rest of the parameters taken from Table
\ref{tab:parameters}. A plant is considered extinct if it attains a density less
than $10^{-6}$ after time $t=20000$.

\section{Results}

\subsection{Fixed preferences}

System (\ref{eq:ode_ss}) models obligatory mutualism between plants and pollinators.
Plants cannot grow in absence of pollinators and pollinators cannot reproduce without
plants. Thus, the trivial equilibrium at which all three species are absent $(P_{1}=P_{2}=A=0)$
is always locally asymptotically stable \cite{may1976,vandermeer_boucher-jtb78},
because when at low population densities, pollinators cannot provide enough pollination
services to plants that will die and, similarly, when at low densities, plants do
not provide enough nectar to support pollinators.

By setting $dP_{1}/dt=dA/dt=0$ with $P_{1}>0,P_{2}=0,A>0$ in (\ref{eq:ode_pl1_ss})
and (\ref{eq:ode_pol_ss}), non-trivial plant 1--pollinator equilibria are

\begin{align}
P_{1\pm} & =\frac{b_{1}e_{1}K_{1}(a_{1}r_{1}-m_{1})u_{1}+dr_{1}w_{1}\pm\sqrt{D_{1}}}{2a_{1}b_{1}e_{1}r_{1}u_{1}}\nonumber \\
A_{1\pm} & =\frac{b_{1}e_{1}K_{1}(a_{1}r_{1}-m_{1})u_{1}-dr_{1}w_{1}\pm\sqrt{D_{1}}}{2b_{1}dr_{1}u_{1}},\label{eq:resideq}
\end{align}

\noindent where $D_{1}=-4b_{1}de_{1}K_{1}m_{1}r_{1}u_{1}w_{1}+(b_{1}e_{1}K_{1}(m_{1}-a_{1}r_{1})u_{1}+dr_{1}w_{1})^{2}$.
These two equilibria are feasible (positive) if $a_{1}r_{1}>m_{1}$ and $D_{1}>0$.
The first is a growth requirement: if not met, even an infinite number of specialized
pollinators (with $u_{1}=1$) cannot prevent plant 1 extinction. The second condition
is met when pollinator preference for plant 1 $(u_{1})$ is above a critical value

\begin{equation}
u_{1a}=\frac{dr_{1}w_{1}}{b_{1}e_{1}(\sqrt{a_{1}r_{1}}-\sqrt{m_{1}})^{2}K_{1}}.\label{eq:u1a}
\end{equation}

By symmetry, there are two non-trivial plant 2--pollinator equilibria $(P_{2\pm},A_{2\pm})$.
They are feasible if $a_{2}r_{2}>m_{2}$ and $D_{2}>0$ ($D_{2}$ is like $D_{1}$
with interchanged sub-indices). The second condition is met when pollinator preferences
for plant 2 are strong enough (i.e., preferences for plant 1 are weak enough) so
that $u_{1}$ is below a critical value $u_{1b}$

\begin{equation}
u_{1b}=1-\frac{dr_{2}w_{2}}{b_{2}e_{2}(\sqrt{a_{2}r_{2}}-\sqrt{m_{2}})^{2}K_{2}}.\label{eq:u1b}
\end{equation}

\noindent In both cases the equilibrium that is closer to the origin ($(P_{1-},A_{1-})$
when plant 2 is missing and $(P_{2-},A_{2-})$ when plant 1 is missing) is unstable.
This instability indicates critical threshold densities. When plant \emph{i} and
pollinator densities are above these thresholds, coexistence is possible. Otherwise,
the system converges on the extinction equilibrium mentioned before. This is a mutualistic
Allee effect \cite{courchampberecgascoigne2008,bronstein2015}.

The equilibrium that is farther from the origin ($(P_{1+},A_{1+})$ when plant 2
is missing and $(P_{2+},A_{2+})$ when plant 1 is missing) will be called the resident
equilibrium. Resident equilibria are stable with respect to small changes in resident
plant and pollinator densities, but may be unstable against invasion of small densities
of the missing plant species. In the case of the plant 1--pollinator equilibrium
$(P_{1+},A_{1+})$, plant 2 invades (i.e., achieves a positive growth rate when rare)
if the competitive effect of plant 1 on plant 2 $(c_{1})$, is smaller than

\begin{equation}
\alpha(u_{1})=\frac{a_{1}b_{1}r_{1}u_{1}K_{2}\left(2b_{2}u_{2}e_{1}K_{1}m_{1}w_{1}(a_{2}r_{2}-m_{2})-m_{2}w_{2}\left(b_{1}e_{1}K_{1}u_{1}(a_{1}r_{1}-m_{1})-dr_{1}w_{1}-\sqrt{D_{1}}\right)\right)}{a_{2}b_{2}r_{2}u_{2}K_{1}m_{1}w_{1}\left(b_{1}e_{1}K_{1}u_{1}(a_{1}r_{1}-m_{1})+dr_{1}w_{1}+\sqrt{D_{1}}\right)},\label{eq:alpha}
\end{equation}

\noindent whereas plant 1 invades the plant 2--pollinator equilibrium $(P_{2+},A_{2+})$
if the competitive effect of plant 2 on plant 1 $(c_{2})$, is smaller than

\begin{equation}
\beta(u_{1})=\frac{a_{2}b_{2}r_{2}u_{2}K_{1}\left(2b_{1}u_{1}e_{2}K_{2}m_{2}w_{2}(a_{1}r_{1}-m_{1})-m_{1}w_{1}\left(b_{2}e_{2}K_{2}u_{2}(a_{2}r_{2}-m_{2})-dr_{2}w_{2}-\sqrt{D_{2}}\right)\right)}{a_{1}b_{1}r_{1}u_{1}K_{2}m_{2}w_{2}\left(b_{2}e_{2}K_{2}u_{2}(a_{2}r_{2}-m_{2})+dr_{2}w_{2}+\sqrt{D_{2}}\right)}.\label{eq:beta}
\end{equation}

\begin{figure}
\begin{centering}
\includegraphics[height=0.9\textheight]{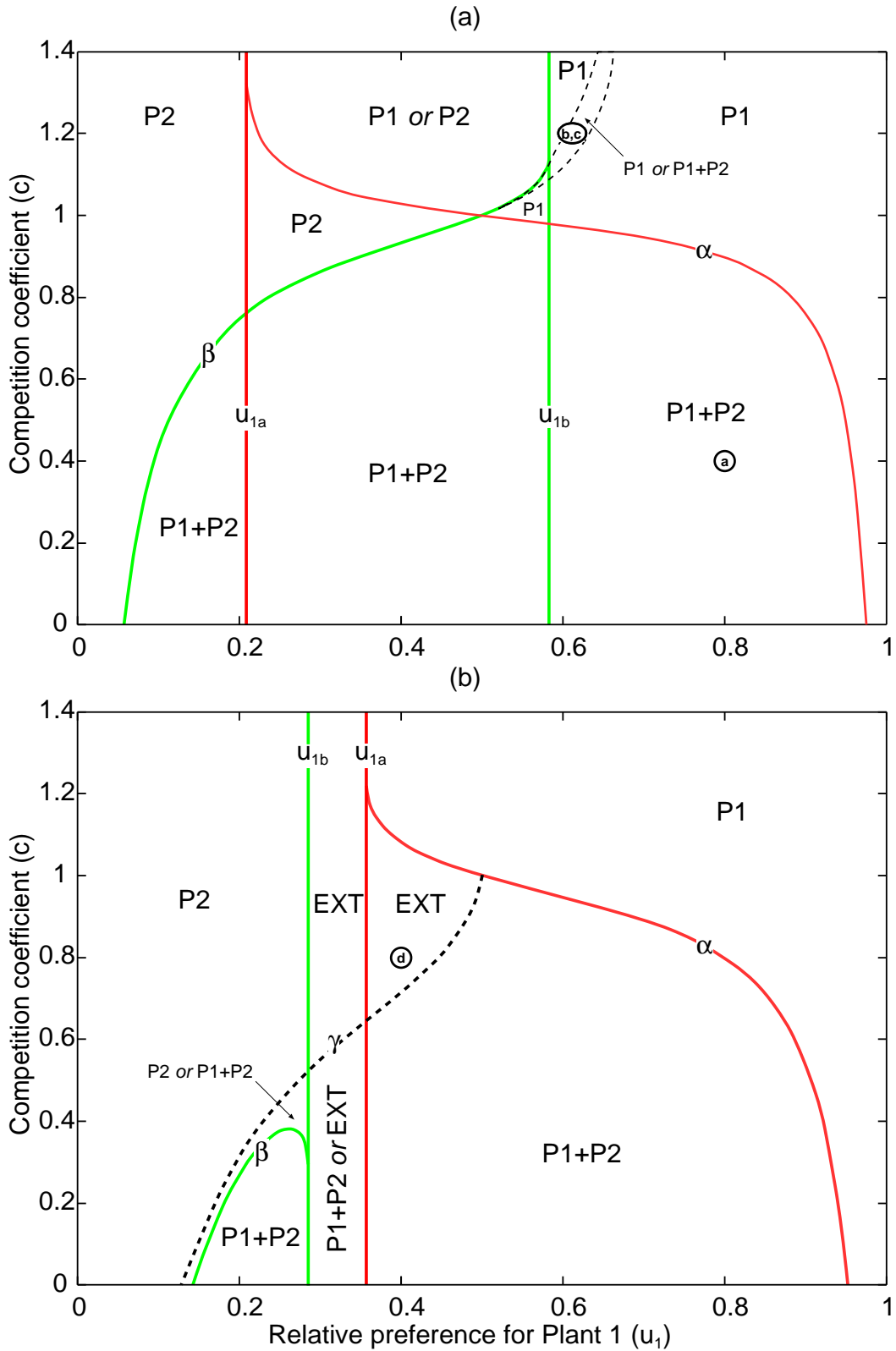} 
\par\end{centering}

\protect\caption{\label{fig:coexistence_parspace}\textbf{Interaction outcomes as a function of competition
strength and fixed pollinator preferences.} Solid lines (coloured, found analytically)
determine regions where single plant equilibria exist and whether they can be invaded
or not. Dashed lines (in black, found numerically, like $\gamma$) determine outcomes
that cannot be predicted by invasibility analysis. Plant 2 can invade Plant 1 in
the region between the red vertical line $u_{1a}$ and the red curve $\alpha$. Plant
1 can invade plant 2 in the region between the green curve $\beta$ and the green
vertical line $u_{1b}$. The final composition of the community is indicated by P1=plant
1 wins, P2=plant 2 wins, P1+P2=coexistence, EXT=extinction of all species; the ``\emph{or}''
separator indicates that the outcome depends on the initial conditions. Parameters
from Table \ref{tab:parameters}, with (a) $K_{i}=60$ and (b) $K_{i}=35$ (i.e.,
above and below critical $K^{*}=37.5$, see (\ref{eq:k-star})). Representative dynamics
for parameter combinations at \textcircled{a}, \textcircled{b}, \textcircled{c}
and \textcircled{d} are illustrated in corresponding panels of Fig \ref{fig:dynamics_fixed}.}
\end{figure}

Functions $\alpha(u_{1})$ and $\beta(u_{1})$ are real when $D_{1}>0$ and $D_{2}>0$,
respectively. In other words, invasibility only makes sense when the plant 1--pollinator
resident equilibrium exists $(u_{1}>u_{1a})$ or, when the plant 2--pollinator resident
equilibrium exists $(u_{1}<u_{1b})$, respectively. The graphs of (\ref{eq:u1a}),
(\ref{eq:u1b}), (\ref{eq:alpha}) and (\ref{eq:beta}) divide the pollinator preference--competition
parameter space into several regions (Fig \ref{fig:coexistence_parspace} where $c=c_{1}=c_{2}$).
Notice that because $\alpha$ and $\beta$ are only feasible to the right of $u_{1a}$
and to the left of $u_{1b}$, respectively, their graphs may or may not intersect
depending on the position of $u_{1a}$ and $u_{1b}$ (cf. panel a vs. b). We show
this by setting a common plant carrying capacity $K=K_{1}=K_{2}$ and making it larger
or smaller than a critical value (Supporting Information \ref{sec:appendixA} Appendix)

\begin{equation}
K^{*}=\frac{(b_{1}e_{1}r_{2}w_{2}(\sqrt{a_{1}r_{1}}-\sqrt{m_{1}})^{2}+b_{2}e_{2}r_{1}w_{1}(\sqrt{a_{2}r_{2}}-\sqrt{m_{2}})^{2})d}{b_{1}b_{2}e_{1}e_{2}(\sqrt{a_{1}r_{1}}-\sqrt{m_{1}})^{2}(\sqrt{a_{2}r_{2}}-\sqrt{m_{2}})^{2}}.\label{eq:k-star}
\end{equation}

Productive environments $(K>K^{*})$ support coexistence of both plant--pollinator
resident equilibria for intermediate pollinator preferences. This is not so in unproductive
environments $(K<K^{*})$, where resident equilibria occur within separated ranges
of pollinator preferences (see below).

\begin{figure}
\begin{centering}
\includegraphics[width=1\textwidth]{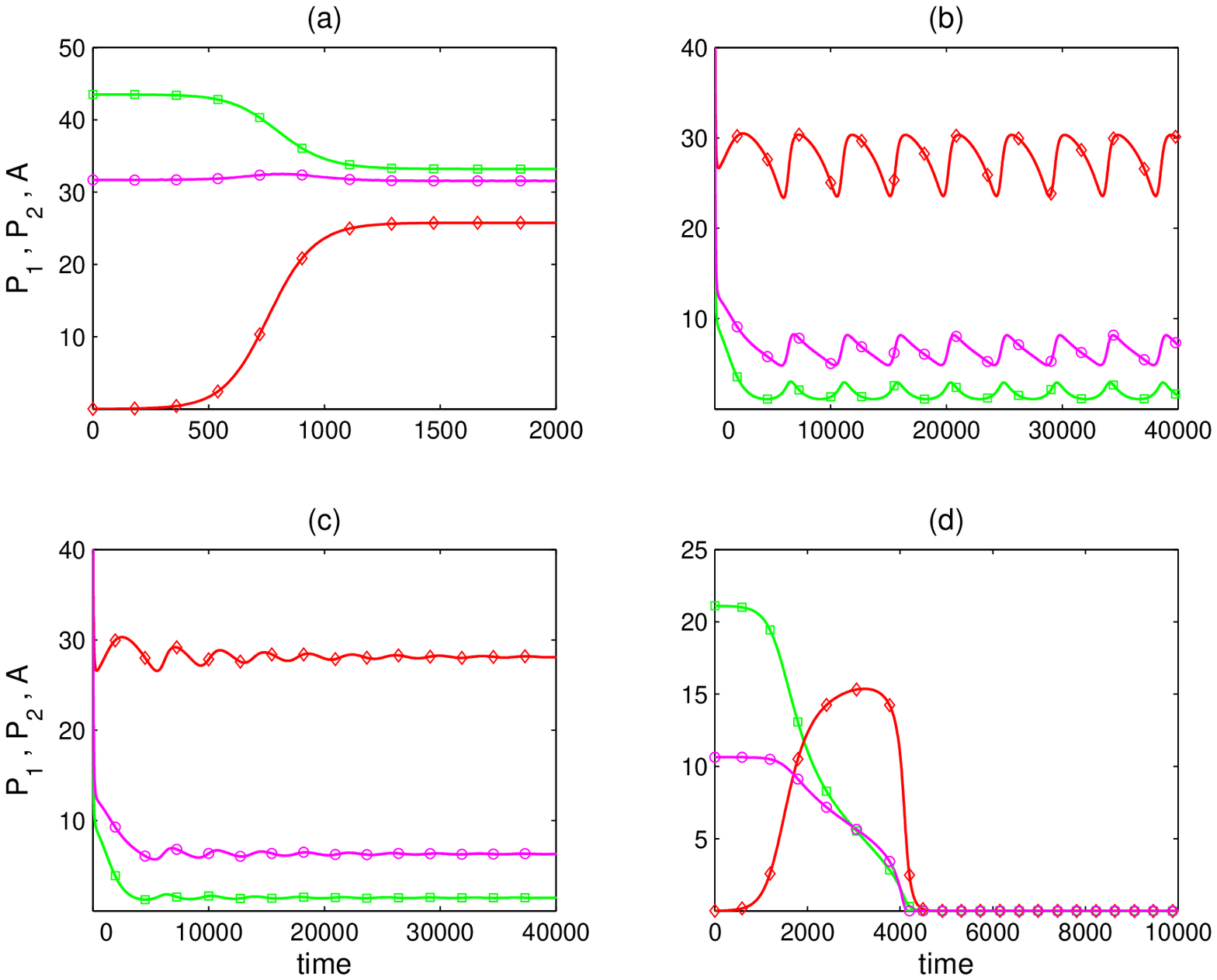} 
\par\end{centering}

\protect\caption{\label{fig:dynamics_fixed}\textbf{Model (\ref{eq:ode_ss}) dynamics with fixed pollinator
preferences.} Population densities are represented by: green squares=plant 1, red
diamonds=plant 2 and pink circles=pollinator. Panels (a) $K_{i}=60$, $u_{1}=0.8$,
$c_{i}=0.4$, (b) $K_{i}=60$, $c_{i}=1.2$, $u_{1}=0.605$, (c) $K_{i}=60$, $c_{i}=1.2$,
$u_{1}=0.607$ and (d) $K_{i}=35$, $u_{1}=0.4$, $c_{i}=0.8$ correspond to positions
of points \textcircled{a}, \textcircled{b}, \textcircled{c} and \textcircled{d}
in Fig \ref{fig:coexistence_parspace}. Other parameters as in Table \ref{tab:parameters}.}
\end{figure}

First, we assume a high plant carrying capacity satisfying $K>K^{*}$. Then $u_{1a}<u_{1b}$,
and $\alpha(u_{1})$ and $\beta(u_{1})$ intersect like in Fig \ref{fig:coexistence_parspace}a.
This leads to several plant invasion scenarios. We start with preferences satisfying
$u_{1a}<u_{1}<u_{1b}$. Such intermediate pollinator preferences allow each species
to coexist with the pollinator at a stable equilibrium. If competition is weak enough
(see the region denoted as ``P1+P2'' in Fig \ref{fig:coexistence_parspace}a),
the missing plant can invade the resident plant--pollinator equilibrium which leads
to both plant coexistence. In contrast, if competition is strong enough (see the
region denoted as ``P1 \emph{or} P2'' in Fig \ref{fig:coexistence_parspace}a),
the missing plant cannot invade. Thus, either plant 1 or plant 2 wins depending on
the initial conditions (i.e., the resident plant that establishes first wins). In
between these two outcomes of mutual invasion and mutual exclusion, there are two
wedge-shaped regions (see regions denoted as ``P1'', and ``P2'' in Fig \ref{fig:coexistence_parspace}a).
In the right (left) region plant 1 (plant 2) invades and replaces plant 2 (plant
1) but not the other way around. The outcomes in the regions of Fig \ref{fig:coexistence_parspace}a
that are either to the left of $u_{1a}$, or to the right of $u_{1b}$ are very different,
because whether the missing plant can invade or not when rare depends entirely on
facilitation by the resident plant. Indeed, let us consider the region of the parameter
space in Fig \ref{fig:coexistence_parspace}a to the right of the vertical line $u_{1b}$
and below the curve $\alpha$. In this region (denoted by ``P1+P2'') pollinator
preference for plant 2 is so low that plant 2 alone cannot support pollinators at
a positive density. It is only due to presence of plant 1 that allows plant 2 survival
through facilitation (Fig \ref{fig:dynamics_fixed}a). Indeed, when plant 1 is resident,
it increases pollinator densities to such levels that allow plant 2 to invade. In
other words, plant facilitation due to shared pollinators widens plant niche measured
as the range of pollinator preferences at which the plant can survive at positive
densities. When inter-specific plant competition is too high (see the region above
the curve $\alpha$ and to the right of $u_{1b}$) plant 2 cannot invade. Similarly,
when pollinator preferences for plant 1 are too low (i.e., to the left of the line
$u_{1a}$), coexistence relies on facilitation provided by plant 2 (resident) to
plant 1 (invader) and on plant competition being not too strong (below curve $\beta$);
if competition is too strong (above $\beta$) plant 1 cannot invade.

Second, we assume low plant carrying capacity satisfying $K<K^{*}$. Then $u_{1a}>u_{1b}$,
and $\alpha(u_{1})$ and $\beta(u_{1})$ never intersect (Fig \ref{fig:coexistence_parspace}b)
in the positive quadrant of the parameter space. For intermediate pollinator preferences
satisfying $u_{1b}<u_{1}<u_{1a}$ coexistence by invasion of the rare plant is not
possible. The reason is that in this region neither plant 1, nor plant 2, can coexist
with pollinators. So, there is no resident system consisting of one plant and pollinators
that could be invaded by the missing rare plant. In regions to the left of $u_{1b}$
plant 2 coexists with pollinators and to the right of $u_{1a}$ plant 1 coexists
with pollinators at a stable equilibrium and invasion conditions for the missing
plant when rare are similar to the case where $K>K^{*}$. Once again, in these regions
coexistence of both plants can be achieved because of the resident plant facilitates
the other plant invasion. The important prediction of this invasion analysis is that
the density mediated indirect interaction between plants through the shared pollinator,
i.e., plant facilitation, increases the set of parameter values for which coexistence
of both plants is possible.

Although invasion analysis proves to be very useful when analysing model (\ref{eq:ode_ss}),
it does not answer the question whether there are some other attractors that cannot
be reached by invasion of the missing species when rare. Using numerical bifurcation
software (XPPAUT \cite{xpp-book}), we found additional outcomes not predicted by
invasibility analysis. When $K>K^{*}$ (Fig \ref{fig:coexistence_parspace}a) invasibility
analysis predicts that plant 2 cannot grow when rare for strong inter-specific plant
competition when $c>\alpha$. However, our numerical analysis shows that it is still
possible for plant 2 to invade provided its initial population density is large enough.
The community dynamics then either oscillate along a limit cycle (Fig \ref{fig:dynamics_fixed}b),
or converge to a stable equilibrium (Fig \ref{fig:dynamics_fixed}c). Such behaviour
was observed in the region denoted by ``P1 \emph{or} P1+P2'' of Fig \ref{fig:coexistence_parspace}a.
This shows that model (\ref{eq:ode_ss}) has multiple attractors (including a limit
cycle). The right dashed boundary of that region corresponds to a fold bifurcation
where a locally stable interior equilibrium merges with an unstable equilibrium and
disappears for higher values of $u_{1}$. Between the two dashed curves there is
another Hopf bifurcation curve (not shown in Fig \ref{fig:coexistence_parspace}a)
where the interior equilibrium looses its stability and a limit cycle emerges. As
preference for plant 1 decreases towards the left dashed boundary, the amplitude
of the limit cycle tends to infinity.

When $K<K^{*}$ we found a curve $\gamma(u_{1})$ that further divides the parameter
space (Fig \ref{fig:coexistence_parspace}b). For the intermediate pollinator preferences
($u_{1b}<u_{1}<u_{1a}$) where neither plant can be a resident, and below $\gamma$
curve (``P1+P2 or EXT''), coexistence is achievable if both plants and the pollinator
are initially at high enough densities. This is an extreme example of plant facilitation.
If combined plant abundances are not large enough, then both plants and the pollinator
go extinct as already predicted by the invasion analysis. Also, if one plant species
is suddenly removed, extinction of pollinator and the other plant follows. Above
the $\gamma$ curve, plant competition is too strong to allow any coexistence and
the outcome is always global extinction (``EXT''). When preference for plant 1
is low $(u_{1}<u_{1b})$, the $\gamma$ curve is slightly above the $\beta$ curve
so that the possible coexistence region is slightly larger than the coexistence region
obtained by invasion of the rare plant (``P2 \emph{or} P1+P2''). However, plant
coexistence in the region between the two curves depends on the initial density of
plant 1: if $P_{1}(0)$ is very low, plant 2 wins as predicted by the invasion analysis,
but if $P_{1}(0)$ is large enough, plant 1 will invade and coexist at an interior
equilibrium with plant 2. In the opposite situation, where preference for plant 1
is very high $(u_{1}>u_{1a})$, $\gamma$ divides the region where plant 2 can invade
(assuming $c<\alpha$) as follows. Below $\gamma$, competition is weak and plant
2 invasion is followed by stable coexistence thanks to resident facilitation. Above
$\gamma$, competition is strong and plant 2 invasion causes plant 1 extinction followed
by plant 2 extinction. This is because pollinator preference for plant 1 is too strong
$(u_{1}>u_{1b})$ which does not allow pollinators to survive on plant 2. Thus, invasion
by plant 2 leads to global extinction (``EXT''). Fig \ref{fig:dynamics_fixed}d
shows an example of such global extinction caused by invasion. Once again, invasion
of plant 2 is possible due to facilitation by plant 1. As plant 2 invades, it has
also an indirect positive effect on plant 1 through facilitation. But this positive
effect does not outweigh the direct negative effect plant 2 has on plant 1 due to
direct competition for resources. Apart from this case of global extinction caused
by invasion, numerical analysis with parameters from Table \ref{tab:parameters}
confirms predictions of our invasion analysis that in the case where one or both
equilibria with one plant missing exist(s), the invasibility conditions $c<\alpha$
and $c<\beta$ predict the existence of a locally stable interior equilibrium at
which both plants coexist with the pollinator.

\subsection{Adaptive preferences}

\subsubsection*{Evolutionarily stable strategy and time scale}

We calculate the evolutionarily stable strategy for fitness defined by (\ref{eq:fitness}).
At the interior (i.e., generalist) behavioural equilibrium the two pay-offs (\ref{eq:payoffs})
must be the same, i.e., $V_{1}=V_{2}$, which yields

\begin{equation}
u_{1}^{*}(P_{1},P_{2},A)=\frac{e_{1}a_{1}P_{1}}{e_{1}a_{1}P_{1}+e_{2}a_{2}P_{2}}+\frac{w_{2}e_{1}a_{1}b_{1}P_{1}-w_{1}e_{2}a_{2}b_{2}P_{2}}{b_{1}b_{2}(e_{1}a_{1}P_{1}+e_{2}a_{2}P_{2})A},\label{eq:u_ees}
\end{equation}

\noindent provided $u_{1}^{*}$ is between 0 and 1. If $V_{1}(u_{1})>V_{2}(u_{1})$
for all $u_{1}$, the ESS is $u_{1}^{*}=1$ and if $V_{1}(u_{1})<V_{2}(u_{1})$ for
all $u_{1}$, the ESS is $u_{1}^{*}=0$. Because

\[
W(u_{1}^{*},u_{1})-W(u_{1},u_{1})=\frac{(Ab_{1}b_{2}(a_{2}e_{2}P_{2}u_{1}-a_{1}e_{1}P_{1}(1-u_{1}))+a_{2}b_{2}e_{2}P_{2}w_{1}-a_{1}b_{1}e_{1}P_{1}w_{2})^{2}}{Ab_{1}b_{2}(a_{1}e_{1}P_{1}+a_{2}e_{2}P_{2})(Ab_{1}u_{1}+w_{1})(Ab_{2}(1-u_{1})+w_{2})}>0
\]

\noindent the interior strategy $u_{1}^{*}$ is also resistant to mutant invasions
\cite{maynardsmith_price-nature73}, i.e., $W(u_{1}^{*},u_{1})>W(u_{1},u_{1})$ for
all strategies $u_{1}\ne u_{1}^{*}$. Thus $u_{1}^{*}$ is the ESS \cite{hofbauersigmund1998}.
We remark that in the ecological literature such an ESS strategy has also been called
the Ideal Free Distribution \cite{fretwell_lucas-actabiot69,krivan_etal-tpb08}.

\begin{figure}
\begin{centering}
\includegraphics[width=0.6\textwidth]{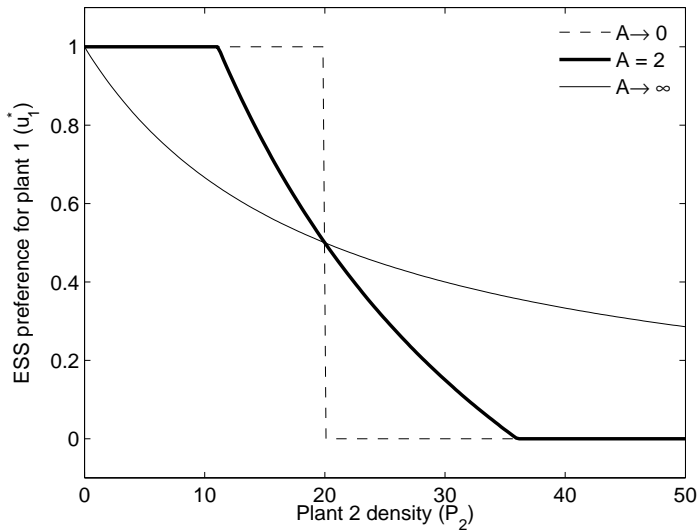} 
\par\end{centering}

\protect\caption{\label{fig:u1ess_vs_p2}\textbf{Evolutionarily stable preference for plant 1 as plant
2 density increases.} Plant 1 density is fixed at $P_{1}=10$ in (\ref{eq:u_ees}).
At very low pollinator density preference switches abruptly (dashed line) from 1
to 0. At very high pollinator density the decline is continuous (thin line). Intermediate
pollinator densities (thick line) cause a combined pattern with switching between
specialisation (horizontal segments) and generalism (decreasing segment). Values
of parameters are those given in Table \ref{tab:parameters}.}
\end{figure}

It follows from (\ref{eq:u_ees}) that as pollinator densities increase, $u_{1}^{*}$
tends to $e_{1}a_{1}P_{1}/(e_{1}a_{1}P_{1}+e_{2}a_{2}P_{2})$, i.e., pollinators
tend towards generalism, with relative preferences reflecting differences in resource
supply rates and quality. This is because at higher pollinator densities fitness
decreases due to intra-specific competition among pollinators for plant resources,
which is compensated for by interacting with the less profitable plant. In contrast,
when pollinator densities become very low, $u_{1}^{*}$ as a function of plant 2
density approximates a step function (Fig \ref{fig:u1ess_vs_p2}). In this latter
case pollinators specialize either on plant 1 or on plant 2 and the switch between
these two possibilities is very sharp. In this case competition between pollinators
is so weak, that pollinators can afford to ignore the less profitable plant.

Equation (\ref{eq:u_ees}) when combined with population dynamics (\ref{eq:ode_ss})
describes the situation where pollinator preferences instantaneously track population
numbers. This situation corresponds to complete time scale separation between behavioural
and population processes. When the assumption of time scale separation is relaxed,
we show that the rate $\nu$ with which pollinator preferences change in (\ref{eq:replicator})
has important effects on plant coexistence.

\begin{figure}
\begin{centering}
\includegraphics[width=1\textwidth]{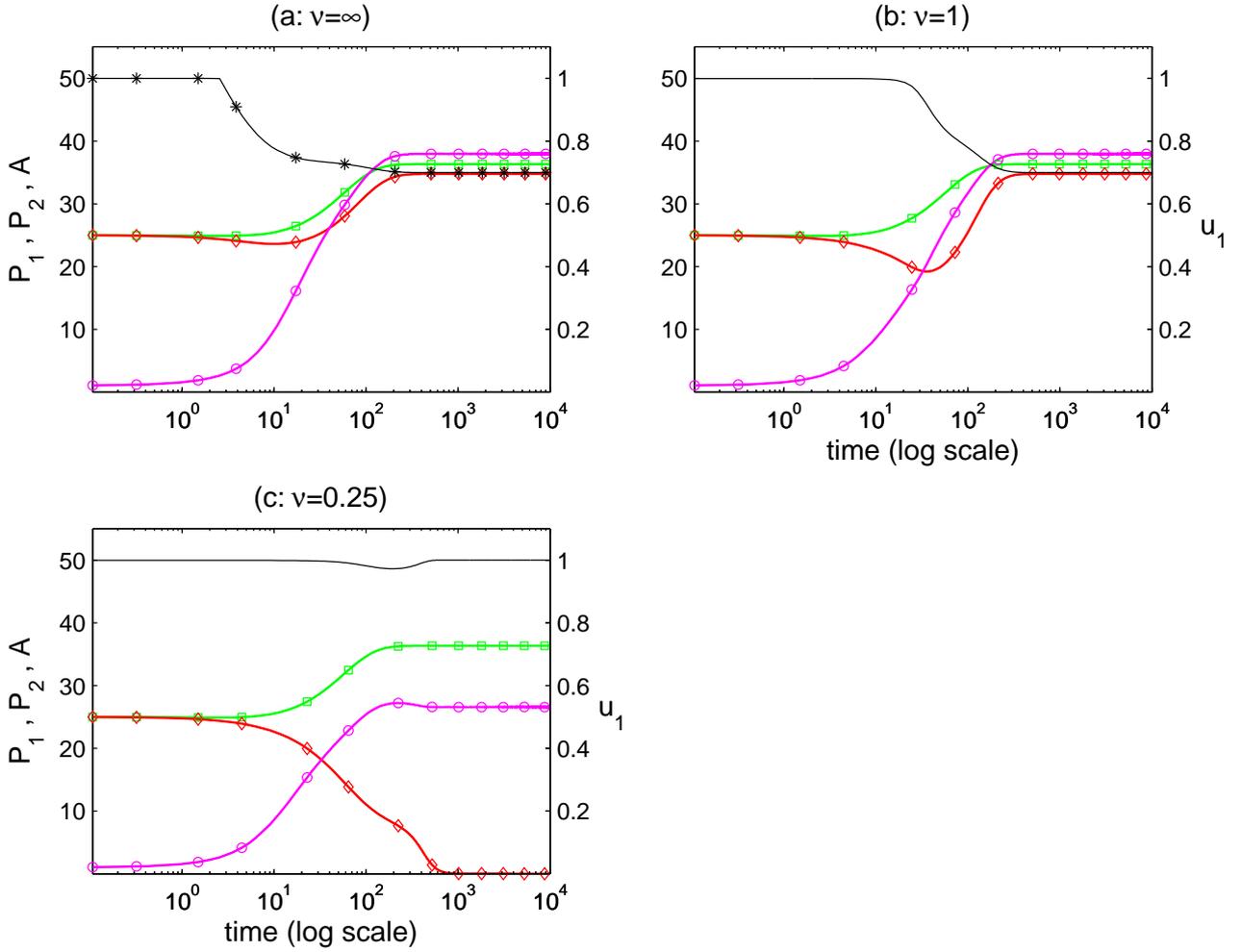} 
\par\end{centering}

\protect\caption{\label{fig:dynamics_flexible}\textbf{Model (\ref{eq:ode_ss}) dynamics with pollinator
adaptive preferences.} Population densities (units in left axes) are represented
by: green squares=plant 1, red diamonds=plant 2, pink circles=pollinator. Pollinator
preference for plant 1 ($u_{1}$, units in right axes) is represented by the black
line. Initial population densities in all panels $P_{1}(0)=P_{2}(0)=25$, $A(0)=1$.
Preferences in (a) are given by the ESS (\ref{eq:u_ees}). Preferences in (b) and
(c) are given by the replicator equation (\ref{eq:replicator}) with $\nu=1$, 0.25
respectively and $u_{1}(0)=0.999$. Parameters as in Table \ref{tab:parameters},
with $K_{i}=50$, $c_{i}=0$.}
\end{figure}

This is especially easy to observe when there is no direct competition between plants
$(c_{1}=c_{2}=0)$. Thus, a plant can cause the decrease of the other plant only
by influencing pollinator preferences. Let us assume that at time $t=0$ both plants
have equal densities and pollinators are rare (but above the critical Allee threshold
density), as shown in Fig \ref{fig:dynamics_flexible}. Because the plant to pollinator
ratio is large, pollinators should specialize on plant 1 $(u_{1}=1)$ which is the
most profitable $(e_{1}>e_{2})$, causing plant 2 to decline and to go extinct, eventually.
However, as pollinator densities start to increase relative to plant densities, pollinators
can become generalists which favours plant coexistence. We start with the assumption
that pollinator preferences track instantaneously population numbers (panel a: $\nu=\infty$),
i.e., $u_{1}=u_{1}^{*}$ is given by the the ESS (\ref{eq:u_ees}) (see the star-line
-$\ast$-$\ast$- in Fig \ref{fig:dynamics_flexible}a). We observe that as pollinator
abundance increases, pollinators become generalists approximately at $t\approx3$,
which is fast enough to prevent plant 2 extinction, and population densities will
tend to an interior equilibrium. When pollinators preference is described by the
replicator equation (panel b: $\nu=1$), we observe that pollinators will become
generalists at a latter time $(t\approx11)$ due to the time lag with which pollinators
preferences follow population abundances. Even with this delay, the decline of plant
2 stops and we obtain convergence to the same population and evolution equilibrium.
However, when adaptation is yet slower, the pollinator preferences will follow changing
population densities with a longer time delay (panel c: $\nu=0.25$), and we get
a qualitatively different outcome with plant 2 extinction. This is because when pollinators
start to behave as generalists $(t\approx100)$, plant 2 abundance is already so
low that it is more profitable for pollinators to switch back to pollinate plant
1 only. We obtain similar results as in Fig \ref{fig:dynamics_flexible} when e.g.,
$c_{1}=c_{2}=0.4$, but coexistence becomes impossible when direct competition becomes
too strong.

In the next section we study combined effects of initial conditions, plant competition
for resources $(c_{i}>0)$, and time scales on plant coexistence.

\subsubsection*{Scenario I (variation of initial plant and pollinator densities)}

\begin{figure}
\begin{centering}
\includegraphics[clip,width=1\textwidth]{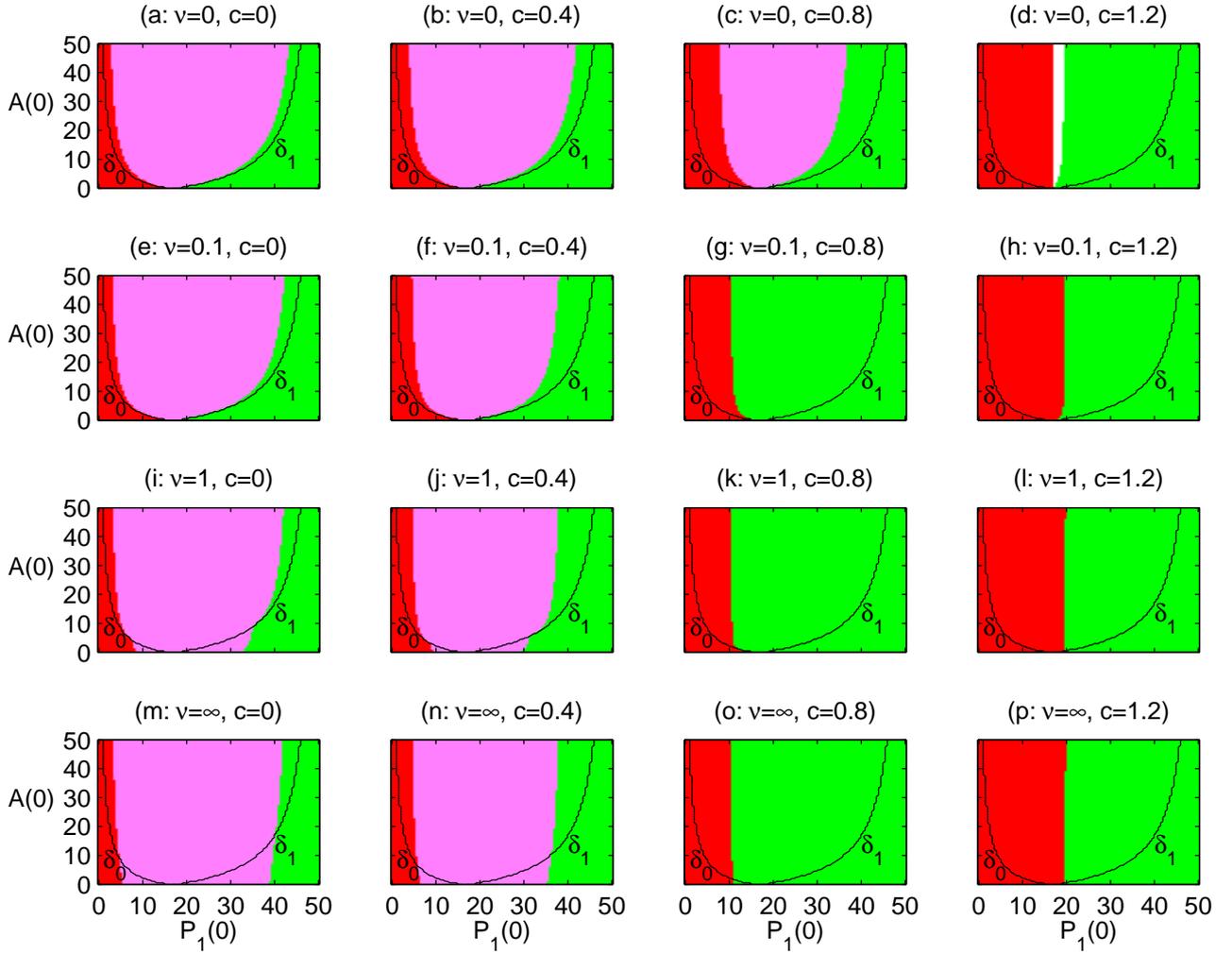} 
\par\end{centering}

\protect\caption{\label{fig:coex_init_plants_pol}\textbf{Effects of foraging adaptation ($\nu$ rows)
and inter-specific competition ($c$ columns) on plant coexistence under scenario
I (variation of initial plant and pollinator densities).} $P_{2}(0)=50-P_{1}(0)$
and $u_{1}(0)$ given by (\ref{eq:u_ees})). Pollinators begin as specialists on
plant 1 to right of line $\delta_{1}$, on plant 2 to the left of line $\delta_{0}$,
and generalists in between. Initial conditions in red and green result in extinction
of plant 1 or 2, respectively. Initial conditions in pink and white result in coexistence
or community extinction, respectively.}
\end{figure}

Here we study the combined effects of population dynamics (\ref{eq:ode_ss}) and
adaptive pollinator preferences (\ref{eq:replicator}) on species coexistence. Fig
\ref{fig:coex_init_plants_pol} shows regions of coexistence (pink), exclusion of
one plant species (red or green), and global extinction (both plants and the pollinator,
white) for different initial plant and pollinator densities. For this scenario combined
initial plant densities are fixed $(P_{1}(0)+P_{2}(0)=50)$. We contrast these predictions
with the situation where population densities are fixed, i.e., when population dynamics
are not considered and pollinator preferences are at the ESS. In this latter case
the necessary condition for both plants to survive is that pollinators behave as
generalists which corresponds to the region between the two curves $\delta_{0}$
and $\delta_{1}$ in Fig \ref{fig:coex_init_plants_pol}. These are the curves along
which the ESS predicts switching between specialist and generalist pollinator behaviour
at initial population densities. These curves are found by solving equation (\ref{eq:u_ees})
for $A$, when $u_{1}^{*}=1$ which yields

\begin{equation}
\delta_{1}\equiv A=\frac{a_{1}e_{1}P_{1}w_{2}}{a_{2}b_{2}e_{2}P_{2}}-\frac{w_{1}}{b_{1}},\label{eq:1-boundary}
\end{equation}
and when $u_{1}^{*}=0$ which yields

\begin{equation}
\delta_{0}\equiv A=\frac{a_{2}e_{2}P_{2}w_{1}}{a_{1}b_{1}e_{1}P_{1}}-\frac{w_{2}}{b_{2}}.\label{eq:0-boundary}
\end{equation}

If population densities do not change, initial conditions to the left (right) of
$\delta_{0}$ $(\delta_{1})$ lead to exclusion of plant 1 (plant 2) because pollinators
specialize on plant 2 (plant 1). Population and pollinator preference dynamics do
change these predictions. The main pattern observed in the simulations is that plant
coexistence becomes less likely as the plant competition coefficient $(c)$ increases.
This is not surprising because a higher inter-specific competition between plants
decreases plant population abundance which makes coexistence of both plants less
likely or impossible (panels c, d, g, h, k, l, o, p). In fact, when inter-specific
plant competition is too strong and the pollinators do not adapt, plant densities
can become so low that the system collapses due to mutualistic Allee effects (panel
d, white region).

The effect of pollinator adaptation rate $(\nu)$ on coexistence is more complex,
in particular when plant competition is moderate or weak (i.e., $c\leq0.4$, panels
a, b, e, f, i, j, m, n). At low adaptation rates ($\nu\leq0.1$, panels a, b, e,
f) increasing the adaptation rate makes the region of coexistence smaller. With faster
adaptation rates (i.e., $\nu\geq1$, panels i, j, m, n), increasing the adaptation
rate further narrows the region of coexistence for large initial pollinator densities,
but widens this region for smaller initial pollinator densities (see the pink areas
below the $\delta_{0}$ and $\delta_{1}$ curves). Although initially pollinators
specialize on the more profitable plant, the inter-specific competition among pollinators
leads to generalism, and provided the adaptation rate is fast enough, to plant coexistence.
This is the same effect as in Fig \ref{fig:dynamics_flexible}, in which the same
set initial conditions with low pollinator densities leads to plant exclusion when
adaptation is slow, or coexistence when adaptation is faster.

When competition is strong $(c=0.8)$, the main effect of pollinator adaptation is
that coexistence entirely disappears (cf. panel c vs. g, k, o). This is because plant
densities are reduced and pollinator densities do not reach high enough densities
that would lead to pollinator generalism. Finally, when competition is very strong
$(c=1.2)$ global extinctions do not happen (cf. panel d vs. h, l p). This is because
adaptation allows pollinators to switch fast enough towards the most profitable plant
before competition drives total plant abundance below the Allee threshold that would
lead to global extinction.

\subsubsection*{Scenario II (variation on initial plant densities and preferences)}

\begin{figure}
\begin{centering}
\includegraphics[clip,width=1\textwidth]{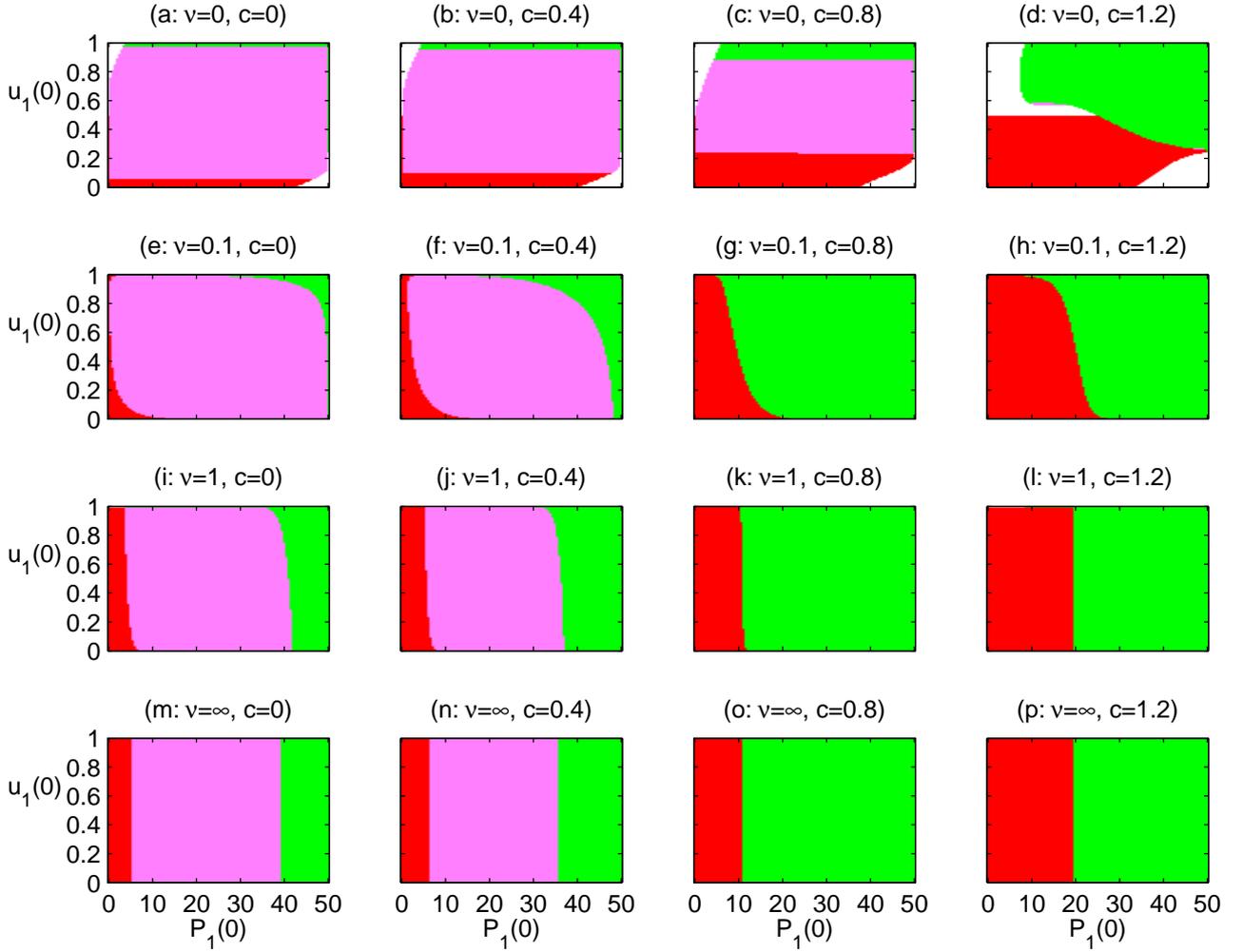} 
\par\end{centering}

\protect\caption{\label{fig:coex_init_plants_pref}\textbf{Effects of foraging adaptation ($\nu$
rows) and inter-specific competition ($c$ columns) on plant coexistence under scenario
II (variation on initial plant densities and preferences).} $P_{2}(0)=50-P_{1}(0)$
and $A(0)=2$). Initial conditions in red and green result in extinction of plant
1 or 2, respectively. Initial conditions in pink and white result in coexistence
or community extinction respectively.}
\end{figure}

This scenario focuses on the effect of initial plant population densities and initial
pollinator preferences on plant coexistence. Similarly to scenario \textbf{I} (Fig
\ref{fig:coex_init_plants_pol}), increasing the inter-specific plant competition
coefficient $c$ makes plant coexistence less likely (Fig \ref{fig:coex_init_plants_pref}).
When pollinators switch from fixed to adaptive foragers the region of plant coexistence
becomes smaller (e.g., see the pink region in the first two columns in Fig \ref{fig:coex_init_plants_pref}).
The general tendency is that increased pollinator adaptation rate reduces the set
of initial conditions that lead to coexistence (cf. third vs. second row in Fig \ref{fig:coex_init_plants_pref}).

For non-adaptive pollinators, community collapse is more widespread than in the scenario
\textbf{I} (cf. white regions in Fig \ref{fig:coex_init_plants_pol} vs. \ref{fig:coex_init_plants_pref}).
This is because in scenario \textbf{II} pollinator preferences can initially be extremely
biased towards the rarest plant (around the upper-left and bottom-right corners of
the panels in Fig \ref{fig:coex_init_plants_pref}). These biased initial preferences
are obviously maladaptive for the pollinator, but in reality, they can be caused
by external disturbances, like the removal of the most preferred plant. For the highest
competition level, communities can collapse when both plants are initially abundant
and pollinators are generalists (around the centre of panel d). This is caused by
the same mechanism outlined for scenario \textbf{I}: plants severely harm each other
for a long time, causing a critical fall in their combined abundance that leads to
extinctions due to the Allee effect. We also observe a very small region where non-adaptive
pollinators can coexist with strongly competing plants (the pink region in panel
d). We examined the corresponding time series to confirm that they display damped
oscillations or limit cycles like in Figs \ref{fig:dynamics_fixed}b,c (results not
shown).

Fast enough pollinator adaptation prevents community collapse, by enabling pollinators
to abandon initially maladaptive diets before it is too late (cf. the first row vs.
e.g., the second row in Fig. \ref{fig:coex_init_plants_pref}). In the long term
either both plants do coexist if plant competition is low (panels e, f, i, j, m,
n), or one plant is excluded by the other plant if plant competition is high (panels
g, h, k, l, o, p). As the adaptation rate increases, Fig \ref{fig:coex_init_plants_pref}
shows an important effect on the general orientation of the regions of coexistence
and exclusion. With no adaptation (top row in Fig \ref{fig:coex_init_plants_pref})
the outcome (coexistence or exclusion) depends more on the initial preferences (vertical
axis) than on the initial plant composition (horizontal axis). However, when adaptation
is fast (e.g., third row $\nu=1$), initial preferences have little influence on
the outcome (unless grossly biased towards 0 or 1) and initial plant composition
is more important. The effect is more sharp when adaptation rate is infinitely fast
($\nu=\infty$), because plant coexistence is entirely independent of the initial
pollinator preferences.

\section{Discussion}

We studied a two-plant--one-pollinator interaction module assuming that pollinator
preferences for plants are either fixed or adaptive. When pollinator preferences
are fixed, we observe that for intermediate pollinator preferences, plants can facilitate
each other indirectly by raising pollinator densities, thus making coexistence more
likely. This effect disappears when preferences are too biased in favour of one plant,
or when competition for factors such as resources or space is too strong. While coexistence
is predominantly at a population equilibrium, coexistence along a limit cycle is
also possible, but under very restrictive conditions in parameter values. Adaptive
pollinator foraging introduces additional competition between plants for pollinators,
because pollinators switch to the major plants, which is bad for rare plants. This
makes coexistence less likely. However, competition for plant resources between pollinators
can promote generalism, thus plant coexistence. The net outcome depends on the relative
speeds between population dynamics and diet adaptation, the strength of competition
between plants for factors other than pollination services, and on the past history
of the community (initial conditions).

\subsection*{Interaction dynamics with fixed preferences}

Under fixed pollinator preferences (i.e., no adaptation) model (\ref{eq:ode_ss})
reveals a rich set of outcomes. The dynamics are complex because plants and pollinators
are obligate mutualists, i.e., their coexistence depends on their population densities
being above the Allee (extinction) threshold. Such thresholds become less important
when one considers alternative pollination mechanisms (e.g., selfing, wind) or mutualistic
partners (other plants, other pollinators), vegetative growth or immigration \cite{vandermeer_boucher-jtb78,bronstein2015},
that our model does not include.

When pollinator preference is extremely biased towards a particular plant, plant
coexistence is not possible even when competition for factors such as space or nutrients
is not considered. This is because the less preferred plant, being rarely pollinated,
cannot increase in abundance. For an intermediate range of pollinator preferences,
coexistence is possible through a number of ways. The most simple and familiar is
coexistence by mutual invasion, like in the Lotka--Volterra competition model. In
this case each plant can attain a positive abundance at an equilibrium with the pollinator
in the absence of the other plant, and the missing plant can invade and establish
in the community (this happens in the middle ``P1+P2'' region of Fig \ref{fig:coexistence_parspace}a).
Another way is when plant \emph{j} can invade the plant \emph{i} and the pollinator
community, but plant \emph{j} alone cannot coexist with the pollinator (left or right
``P1+P2'' regions in Fig \ref{fig:coexistence_parspace}a,b). In all these cases
one plant (\emph{i}) \emph{facilitates} the other plant (\emph{j}) via pollinator
sharing, by increasing the pollinator density (see Fig \ref{fig:model}b). This indirect,
density mediated interaction between plants \cite{bolker_etal-ecology03} is called
pollinator mediated facilitation \cite{rathcke-real1983}. A striking example of
pollinator mediated facilitation occurs when neither plant can coexist with the pollinator
without the other plant, but pollinators do coexist with both plants (this happens
in the ``P1+P2 \emph{or} EXT'' region in Fig \ref{fig:coexistence_parspace}b).
However, for trajectories to converge to the interior equilibrium the initial plant
densities must be high enough so that the pollinator mediated facilitation is strong
enough. Pollinator mediated facilitation \cite{rathcke-real1983} has been empirically
documented \cite{moeller-ecology04,ghazoul-joe06}, and its role in plant invasions
recognized \cite{morales_traveset-ecolett09}.

Invading plants can have positive or negative effects on the resident species. If
plant competition is very weak or absent $(c_{i}\approx0)$, the invader can indirectly
increase the resident's plant density. This is another manifestation of pollinator
mediated facilitation \cite{rathcke-real1983}, this time by the invader. If competition
is stronger, invasion and establishment can cause decline in the resident plant (e.g.,
Fig. \ref{fig:dynamics_fixed}a) or its replacement by the invader, as expected according
to competition theory \cite{grover1997}. In this case plant competition just outweighs
facilitation. Our analysis also shows that in low productive environments (i.e.,
when (\ref{eq:k-star}) does not hold), a resident plant can facilitate the invasion
of poor quality plants (with low $e_{i}$) that cause the subsequent collapse of
the whole community (e.g., Fig. \ref{fig:dynamics_fixed}d).

Numerical analysis of model (\ref{eq:ode_ss}) shows that when coexistence cannot
occur by invasion when rare, it is sometimes achievable if the invader's density
is initially large enough. Coexistence generally takes place at stable densities
(e.g., region ``P2 \emph{or} P1+P2'' in Fig \ref{fig:coexistence_parspace}b and
most of region ``P1 \emph{or} P1+P2'' in Fig \ref{fig:coexistence_parspace}a).
But we also found coexistence along a limit cycle. Limit cycles occur for very narrow
ranges of preferences and under strong competition (e.g., a small part of region
``P1 \emph{or} P1+P2'' in Fig \ref{fig:coexistence_parspace}a). We only found
limit cycles when low quality plant 2 ($e_{2}<e_{1}$, Table \ref{tab:parameters})
cannot support the pollinator and cannot invade plant 1--pollinator equilibrium.
Only when plant 2 enters at large densities, it will start driving out plant 1, followed
by the pollinator. This leads to plant 2 decline and the later recovery of the plant
1--pollinator system, completing the cycle. We could say that such dynamics between
plant 1--pollinator subsystem and plant 2 resembles prey--predator or host--parasite
interactions. Limit cycles in competitor--competitor--mutualist modules have been
predicted before, in models of the Lotka--Volterra type \cite{gyllenberg_etal-physicad06}.
We never observed limit cycles when pollinator preferences are adaptive ($\nu>0$
in equation (\ref{eq:replicator})).

We assumed that plant competition affects growth rather than death rates \cite{benadi_etal-amnat13,valdovinos_etal-oikos13}.
This assumption is sound when plants are mainly limited by space, or by resources
whose access are linked to space, such as light. In such circumstances a plant could
produce many seeds thanks to pollination, but space puts a limit on how many will
recruit as adults. It remains to see how our results would change if competition
is considered differently, when adult plant mortality is affected by competition
(e.g., \cite{bastolla_etal-nature09}). This can be very important under scenarios
of interference like allelopathy or apparent competition caused by herbivores \cite{holt-tpb77}.

\subsection*{Adaptive preferences and population feedbacks}

The ESS (\ref{eq:u_ees}) predicts that when at low densities, pollinators will pollinate
only the plant that is most profitable, while at higher densities they will tend
to pollinate both plants. This positive relationship between pollinator/consumer
abundance and generalism was experimentally demonstrated for bumblebees \cite{fontaine_etal-joe08}.

Plant and pollinator densities are not static, they change within the limits imposed
by several factors: e.g., space and nutrients in the case of plants, or plant resources
such as nectar in the case of pollinators. On the other hand, plants and pollinators
require minimal critical densities of each other in order to compensate for mortality.
Thus, a given ESS at which one plant is excluded from the pollinator's diet will
cause that plant to decrease in density, and, possibly, to go extinct. However, as
population densities change the ESS can also change in ways that may favour coexistence.
These outcomes will depend on the time scale of pollinator foraging adaptation. For
this reason, we introduced the replicator equation (\ref{eq:replicator}) as a dynamic
description of pollinator preferences and we coupled it with population dynamics.
One of the main consequences of introducing replicator dynamics is the disappearance
of complex dynamics such as limit cycles or global extinctions triggered by invasion
(Fig. \ref{fig:dynamics_fixed}b,d). In contrast, the dynamics with pollinator adaptation
are characterized by fewer stable outcomes (plant 1 only, plant 2 only, coexistence)
with strong dependence on the initial conditions.

Whether or not adaptive pollinator preferences promote plant coexistence depends
critically on the strength of competition (i.e., competition coefficient) and the
rate of pollinator adaptation $(\nu)$. In our simulations, we determined the region
of initial population densities and pollinator preferences leading towards plant
coexistence, as a function of these two factors. The larger this region, the more
likely plant coexistence. As competition strength increases, coexistence becomes
less likely as expected from competition theory \cite{grover1997}. As pollinator
adaptation rates increase the pattern is more complex and sometimes equivocal, as
adaptation can increase or decrease the likelihood of coexistence (Figs \ref{fig:coex_init_plants_pol}
and \ref{fig:coex_init_plants_pref}). For example, when competition is weak or moderate
in simulation scenario \textbf{I}, we see that the region of coexistence is generally
wider when pollinator densities are initially large and more narrow when pollinators
are initially rare (Fig. \ref{fig:coex_init_plants_pol} for $c\leq0.4$). This agrees
with the pattern outlined in Figs \ref{fig:model}b and c. In other words, when pollinators
are abundant competition between pollinators promotes generalism, which is good for
plant coexistence, whereas if pollinators are rare they can easily turn into specialists,
which is bad for coexistence. However, when the adaptation rate increases, we also
observe that if pollinators are initially rare, the region of coexistence widens.
Fig \ref{fig:dynamics_flexible} can help explain this: a rare pollinator specializes
on a single plant, even when neither plant is too rare (e.g., initial ESS panel a).
As pollinators start growing, competition for plant resources will cause pollinators
to drive towards generalism. If this change is fast enough (large $\nu$, panel b)
the extinction of the less preferred plant can be prevented. However, if the change
is too slow (small $\nu$, panel c) the less preferred plant declines too fast to
be rescued from extinction. Thus, the results from scenario \textbf{I} tells us that
time lags in pollinator adaptation with respect to population dynamics have important
consequences for plant coexistence.

Simulation scenario \textbf{II} also tells us that adaptation lags can affect the
entire community, plants and pollinators. In this scenario initial pollinator preferences
are arbitrary (i.e., not at the ESS). This is likely to happen if external perturbations
(e.g., disease, grazing) makes the most preferred plant too rare and the less preferred
too common, in a very short time. If pollinators cannot turn into generalists fast
enough they will go extinct because of the mutualistic Allee effect. This leads to
the collapse of the community (white regions in Fig. \ref{fig:coex_init_plants_pref}
at $\nu=0$). When pollinators are able to adapt, such global extinctions can be
prevented, sometimes at the price of one plant going extinct. We also observe global
extinction in scenario \textbf{II} when initial plant ratios and fixed pollinator
preferences are both not too biased (i.e., around the centre of panel d in Fig. \ref{fig:coex_init_plants_pref}).
In these particular cases, generalism is not optimal because of splitting foraging
effort on both plants, neither plant gets enough pollination services to survive.
By adapting its preference towards a single plant (panels h, l, p in Fig. \ref{fig:coex_init_plants_pref}),
the pollinator population would avoid extinction.

Adaptive pollination in a mutualistic interaction module predicts opposite trends
for biodiversity when compared with the apparent competition food web module with
adaptive consumers. Instead of promoting species coexistence by decreasing competitive
asymmetries as in the apparent competition food web module \cite{krivan-amnat97,kondoh-science03,krivan-jtb14},
adaptive pollinator preferences can increase or decrease plant competitive asymmetries,
making their coexistence less or more likely, respectively. At low density, pollinators
tend to specialize on the most common plant (Fig \ref{fig:model}c), leading to the
exclusion of the rare plant. At high density, competition between pollinators promotes
generalism (Fig \ref{fig:model}d), which helps in promoting plant coexistence. Similar
outcomes are predicted in the model studied by Song and Feldman \cite{song_feldman-rspb14},
where plant coexistence is favoured under low plant:pollinator ratios (although these
ratios were kept fixed by these authors). Because plant:pollinator ratios are dynamic,
transitions from specialization to generalism depend not only on adaptation rates,
but also on how fast pollinator densities react to simultaneous changes in plant
densities. We see this in Fig \ref{fig:dynamics_flexible}, where pollinators attain
large density very quickly, before one plant becomes too common and the other too
rare. This indicates that the form of \emph{population regulation} (e.g., linearly
decreasing for plants \cite{benadi_etal-amnat13,valdovinos_etal-oikos13} vs. hyperbolically
decreasing for pollinators \cite{schoener-tpb76}) as well as the \emph{numerical
response} towards mutualistic partners (e.g., saturating for plants vs. linear for
pollinators \cite{revilla-jtb15}), can play important roles in consumer adaptation
in mutualistic communities.

Our simulations assume that plant 1 is richer in energy rewards when compared to
plant 2 $(e_{1}>e_{2})$ while we keep all the other plant-specific parameters equal.
We also ran simulations with plant 1 being better with respect to other plant-specific
parameters (e.g., $r_{1}>r_{2}$ or $w_{2}>w_{1}$, keeping $e_{i}=0.1$ and the
rest of parameters as in Table \ref{tab:parameters}). In these simulations (not
shown here) coexistence is generally more difficult to attain (e.g., coexistence
regions as those in Fig \ref{fig:coex_init_plants_pol} get smaller). The reason
is that in our model, plant rewards $(e_{i})$ affect plants only indirectly, by
influencing pollinator preferences. In contrast, other plant-specific parameters
affect plant dynamics directly. Finally, the larger $c_{i}$ and $K_{i}$ the more
likely plant \emph{i} always win in competition, but this is a natural result expected
in models derived from the Lotka--Volterra competitive equations.

Some predictions from our model are in qualitative agreement with experiments. For
example \cite{ghazoul-joe06} shows transitions from plant facilitation to competition
for pollinators \cite{rathcke-real1983} when one plant species (\emph{Raphanus raphanistrum})
is exposed to increasing numbers of an alternative plant (\emph{Cirsium arvense}).
In the same study, the relative visitation frequency of a plant (\emph{Raphanus})
declines faster than predicted by the decline in the relative proportion of its flowers
\cite{ghazoul-joe06}. The ESS can explain this outcome as the superposition of a
relative resource availability effect and a resource switching effect (i.e., first
and second terms respectively, in the right-hand-side of (\ref{eq:u_ees})), as shown
by Fig \ref{fig:u1ess_vs_p2} (compare it with figure 6 in \cite{ghazoul-joe06}).
The effect of resource competition on the relationship between pollinator density
and generalism, was demonstrated by another experiment \cite{fontaine_etal-joe08}.
Other studies show that invasive plant species can take advantage of changing pollinator
preferences, increasing their chances to get included into native communities \cite{brown_etal-ecology02}.
Finally, one meta-analysis indicates that pollinators can be taken away by invasive
plants, affecting native plants adversely \cite{morales_traveset-ecolett09}.

\subsection*{Inter-specific pollen transfer effects}

Model (\ref{eq:ode_ss}) considers only one single pollinator species. This makes
pollinator generalism (i.e., $u_{1}$ strictly between 0 or 1) a requisite for coexistence.
However, when pollinators are generalists, rare plants would experience decreasing
pollination quality, due to the lack of constancy of individual pollinators delivering
non-specific pollen or losing con-specific pollen \cite{levin_anderson-amnat70,murcia_feinsinger-ecology96,morales_traveset-crps08}.
We do not consider inter-specific pollen transfer effects (IPT) in this article.
Modelling IPT effects requires additional assumptions about visitation probabilities
\cite{valdovinos_etal-oikos13}, pollen carry-over \cite{benadi_etal-amnat13} or
pollinator structure \cite{kunin_iwasa-tpb96,song_feldman-rspb14}. Nevertheless,
we simulated scenarios \textbf{I} and \textbf{II} again, but replacing our equations
(\ref{eq:ode_ss}) by a system of equations that considers IPT \cite{valdovinos_etal-oikos13}.
We found that most of our results hold qualitatively, i.e., the coexistence regions
display the same patterns like in Figs \ref{fig:coex_init_plants_pol} and \ref{fig:coex_init_plants_pref}
(results not shown).

There is no question that IPT affects pollination efficiency. However, the relative
importance of IPT may also depend on the structure of the environment where interactions
occur. A survey of field and laboratory results \cite{morales_traveset-crps08} reports
that in spite of strong IPT effects on plant reproduction for certain systems, many
studies found little or no significant effects in other systems. One reason could
be the scale of the system under study, which can influence the way mobile pollinators
experience the resource landscape: fine grained or coarse grained, e.g., well mixed
or patchy. Thus, if plant species are not totally intermingled, but also not isolated
in clumps, the negative effects of IPT on seed set (a proxy for plant fitness) could
be reduced \cite{thomson-oikos82}. In addition, unless we consider a single flower
per plant at any time, the resource is almost always patchy. This means that IPT
effects in self-compatible plants would be stronger just after pollinator arrival,
decreasing for the remaining flowers before the pollinator leaves the plant.

\subsection*{From modules to networks and from adaptation to co-evolution}

The scope of our work is limited to adaptation in a single pollinator species only.
In real life settings adaptation can be affected by (i) competition among several
pollinator species, and by (ii) plant--pollinator co-evolution.

With respect to point (i), large community simulations \cite{valdovinos_etal-oikos13}
indicate that inter-specific competition can force pollinators to change their preferences
in order to minimise niche overlap. This can promote coexistence and specialization
on rare plants at risk of competitive exclusion. Song and Feldman \cite{song_feldman-rspb14}
discovered a similar mechanism, with a polymorphic pollinator, i.e., consisting of
specialist and generalist sub-populations. Thus, adding a second pollinator would
be a next step to consider, in order to address inter-specific competition.

Addressing point (ii) will require trade-offs in plant traits. We showed how differences
in pollinator efficiencies $(e_{i})$ indirectly affect plant dynamics (\ref{eq:ode_ss}).
However, pollinator efficiencies can depend on plant allocation patterns, which can
affect their growth, mortality or competitive performance $(r_{i},m_{i},c_{i})$.
Plant adaptation likely happens over generations, so a replicator equation approach
or adaptive dynamics \cite{geritz_etal-evoeco98} will be useful to study plant--pollinator
co-evolution.

In spite of the complexity of real plant pollinator networks, small community modules
will remain useful to tease apart the mechanisms that regulate their diversity. Models
of intermediate complexity like (\ref{eq:ode_ss}) can help us discover important
results concerning interaction dynamics, pollinator foraging patterns (e.g., pollinator
ESS) and the consequences of differences between ecological vs. adaptation time scales.

\section*{Acknowledgements}

Support provided by the Institute of Entomology (RVO:60077344) is acknowledged. This
project has received funding from the European Union's Horizon 2020 research and
innovation programme under the Marie Sk{\l }odowska-Curie grant agreement No 690817.

 \bibliographystyle{vancouver}
\bibliography{pollifor}

\singlespacing 

\newpage{}

\appendix
\begin{center}
\textsf{\textbf{\Large{}Supporting Information of}}
\par\end{center}{\Large \par}

\begin{center}
\textsf{\textbf{\Large{}``Pollinator foraging flexibility and coexistence of competing
plants''}}
\par\end{center}{\Large \par}

\begin{center}
\textsf{\textbf{\Large{}Tomás A. Revilla \& Vlastimil K\v{r}ivan}}
\par\end{center}{\Large \par}

\renewcommand{\thesection}{S\arabic{section}}
\setcounter{section}{0}
\setcounter{page}{1}
\renewcommand{\theequation}{S.\arabic{equation}}
\setcounter{equation}{0}
\renewcommand{\thefigure}{S.\arabic{figure}}
\setcounter{figure}{0}
\renewcommand{\thetable}{S.\arabic{table}}
\setcounter{table}{0}

\section{Analysis with fixed preferences}

\label{sec:appendixA}

The community model used in the main text can be derived from a mass action mechanism
that considers plant resource dynamics explicitly%
\footnote{Revilla, T. A. (2015) Numerical responses in resource-based mutualisms: a time scale
approach, Journal of Theoretical Biology, 378:39--46.%
}

\begin{subequations}\label{eq:ode_mechanism} 
\begin{align}
\frac{dF_{i}}{dt} & =a_{i}P_{i}-w_{i}F_{i}-u_{i}b_{i}F_{i}A\label{eq:ode_flowers}\\
\frac{dP_{i}}{dt} & =\left\{ \begin{array}{ll}
r_{i}u_{i}b_{i}F_{i}A\left(1-\frac{P_{i}+c_{j}P_{j}}{K_{i}}\right)-m_{i}P_{i} & \text{if }P_{i}>0\\
0 & \text{if }P_{i}=0
\end{array}\right.\label{eq:ode_plants}\\
\frac{dA}{dt} & =(e_{1}u_{1}b_{1}F_{1}+e_{2}u_{2}b_{2}F_{2}-d)A,\label{eq:ode_pollinators}
\end{align}
\end{subequations}

\noindent where $i,j=1,2$ with $i\neq j$. In this model $F_{i}$ denotes density
of plant $i$ resources such as nectar. Note that pollinator birth rates are directly
proportional to plant resources, like in most consumer--resource models. Plant birth
rates are proportional to the product between plant resource and pollinator densities,
on the assumption that the rate of plant pollination relates linearly with the rate
of pollinator resource consumption. In (\ref{eq:ode_plants}) we prevent plant \emph{i}
to reach negative densities by setting their population growth to zero when there
is no plant \emph{i}. Next, we assume that resources equilibrate quickly with current
plant and pollinator densities (i.e., $dF_{i}/dt=0$, while $dP_{i}/dt\neq0$ and
$dA/dt\neq0$). Thus $F_{i}=a_{i}P_{i}/(w_{i}+u_{i}b_{i}A)$, which we substitute
in (\ref{eq:ode_plants}) and (\ref{eq:ode_pollinators}), to get the system of ordinary
differential equations (ODE) shown in the main text as ``(1)''. Note that the ODE
system in the main text keeps the positive octant invariant (i.e., non-negative)
so we do not need any additional assumption on plant growth when at zero density.
Our analysis is much easier to follow if we re-arrange the ODE in a form that resembles
classical competition (Lotka--Volterra) and consumer--resource models

\begin{align}
\frac{dP_{1}}{dt} & =g_{1}(A)\left(1-\frac{P_{1}+c_{2}P_{2}}{k_{1}(A)}\right)P_{1}\nonumber \\
\frac{dP_{2}}{dt} & =g_{2}(A)\left(1-\frac{P_{2}+c_{1}P_{1}}{k_{2}(A)}\right)P_{2}\label{eq:lvodes}\\
\frac{dA}{dt} & =\left(e_{1}h_{1}(A)P_{1}+e_{2}h_{2}(A)P_{2}-d\right)A\nonumber 
\end{align}

\noindent with

\begin{align}
g_{i}(A) & =\frac{r_{i}a_{i}u_{i}b_{i}A}{w_{i}+u_{i}b_{i}A}-m_{i}\label{eq:r(P)}\\
k_{i}(A) & =K_{i}\left(1-\frac{m_{i}(w_{i}+u_{i}b_{i}A)}{r_{i}a_{i}u_{i}b_{i}A}\right)\label{eq:k(P)}\\
h_{i}(A) & =\frac{a_{i}u_{i}b_{i}}{w_{i}+u_{i}b_{i}A}.\label{eq:h(A)}
\end{align}

\noindent The plant intrinsic growth rates $g_{i}(A)$ and the carrying capacities
$k_{i}(A)$ are saturating functions of pollinator density, i.e., $\lim_{A\to\infty}g_{i}(A)=r_{i}a_{i}-m_{i}$
and $\lim_{A\to\infty}k_{i}(A)=K_{i}(1-m_{i}/(r_{i}a_{i}))$. Pollinator per capita
consumption rates $h_{i}(A)$ decrease to 0 with increasing pollinator density due
to intra-specific pollinator competition for plant resources. We observe that at
low pollinator densities both $g_{i}$ and $k_{i}$ are negative.

System (\ref{eq:lvodes}) has the extinction equilibrium $(P_{1},P_{2},A)=(0,0,0)$.
The Jacobian matrix evaluated at this equilibrium is

\begin{equation}
J(0,0,0)=\left[\begin{array}{ccc}
-m_{1} & 0 & 0\\
0 & -m_{2} & 0\\
0 & 0 & -d
\end{array}\right].\label{eq:jac000}
\end{equation}

Thus, all eigenvalues are negative and the trivial equilibrium is locally asymptotically
stable. There are also other, non-trivial equilibria that we consider next.

\subsection{Single plant--pollinator equilibria}

\begin{figure}
\begin{centering}
\includegraphics{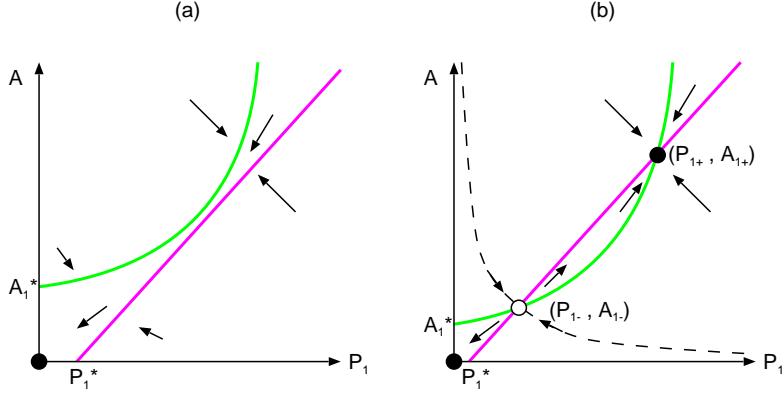} 
\par\end{centering}

\protect\caption{\label{fig:isoclines2d}Plant 1 and pollinator phase plane. The plant and pollinator
(non trivial) nullclines are coloured green and pink respectively. (a) When (\ref{eq:u1a_fold})
does not hold the nullclines don't intersect and thus both species go extinct. (b)
When (\ref{eq:u1a_fold}) holds the nullclines intersect at two equilibrium points:
a saddle point which is unstable (circle) and a locally stable node (dot). Plant
and pollinator coexist for combinations of densities above the separatrix passing
through the saddle point (dash line).}
\end{figure}

Let us assume that plant 2 is absent and we study the plant 1--pollinator subsystem.
By setting $P_{2}=0$ the nullcline of plant 1 is

\begin{equation}
P_{1}=k_{1}(A)=K_{1}\left(1-\frac{m_{1}(w_{1}+u_{1}b_{1}A)}{r_{1}a_{1}u_{1}b_{1}A}\right),\label{eq:p1null}
\end{equation}
see Fig \ref{fig:isoclines2d}. The plant nullcline crosses the $A$ axis at

\begin{equation}
A_{1}^{*}=\frac{m_{1}w_{1}}{u_{1}b_{1}(r_{1}a_{1}-m_{1})}\label{eq:astar}
\end{equation}

\noindent and it has a vertical asymptote at

\begin{equation}
P_{1}=K_{1}\left(1-\frac{m_{1}}{r_{1}a_{1}}\right).\label{eq:nimax}
\end{equation}

The plant 1 nullcline is in the positive quadrant of the plant 1--pollinator phase
space provided

\begin{equation}
r_{1}a_{1}>m_{1}.\label{eq:ra>m}
\end{equation}

\noindent Setting $P_{2}=0$ in (\ref{eq:lvodes}) we get the pollinator nullcline

\noindent 
\begin{equation}
P_{1}=\frac{d}{e_{1}h_{1}(A)}=\frac{d(w_{1}+u_{1}b_{1}A)}{e_{1}a_{1}u_{1}b_{1}}\label{eq:anull}
\end{equation}
which crosses $P_{1}$ axis at

\noindent 
\begin{equation}
P_{1}^{*}=\frac{dw_{1}}{e_{1}a_{1}u_{1}b_{1}}.\label{eq:n1star}
\end{equation}

Fig \ref{fig:isoclines2d} shows two possible nullcline configurations. Provided
preference for plant 1 is strong enough and satisfies

\begin{equation}
u_{1}>u_{1a}=\frac{dr_{1}w_{1}}{b_{1}e_{1}(\sqrt{a_{1}r_{1}}-\sqrt{m_{1}})^{2}K_{1}},\label{eq:u1a_fold}
\end{equation}
the nullclines intersect at two positive equilibria (Panel b) $(P_{1-},A_{1-})$
and $(P_{1+},A_{1+})$ where

\[
P_{1\pm}=\frac{b_{1}e_{1}K_{1}(a_{1}r_{1}-m_{1})u_{1}+dr_{1}w_{1}\pm\sqrt{D_{1}}}{2a_{1}b_{1}e_{1}r_{1}u_{1}}
\]
\[
A_{1\pm}=\frac{b_{1}e_{1}K_{1}(a_{1}r_{1}-m_{1})u_{1}-dr_{1}w_{1}\pm\sqrt{D_{1}}}{2b_{1}dr_{1}u_{1}},
\]
and

\begin{equation}
D_{1}=-4b_{1}de_{1}K_{1}m_{1}r_{1}u_{1}w_{1}+(b_{1}e_{1}K_{1}(m_{1}-a_{1}r_{1})u_{1}+dr_{1}w_{1})^{2}.\label{eq:disc1}
\end{equation}
When $u_{1}$ does not meet the threshold in (\ref{eq:u1a_fold}), no positive interior
equilibrium exists (Panel a).

The Jacobian matrix evaluated at one of these two interior equilibria (i.e., $(P_{1},A)=(P_{1+},A_{1+})$
or $(P_{1},A)=(P_{1-},A_{1-})$) is

\begin{equation}
J(P_{1},A)=\left[\begin{array}{cc}
-\frac{Aa_{1}b_{1}P_{1}r_{1}u_{1}}{K_{1}(Ab_{1}u_{1}+w_{1})} & \frac{a_{1}b_{1}(K_{1}-P_{1})P_{1}r_{1}u_{1}w_{1}}{K_{1}(Ab_{1}u_{1}+w_{1})^{2}}\\
\frac{Aa_{1}b_{1}e_{1}u_{1}}{Ab_{1}u_{1}+w_{1}} & -\frac{Aa_{1}b_{1}^{2}e_{1}P_{1}u_{1}^{2}}{(Ab_{1}u_{1}+w_{1})^{2}}
\end{array}\right].\label{eq:jacnew}
\end{equation}
We observe that the trace of the Jacobian is negative and the determinant is

\[
\det(J)=\frac{Aa_{1}^{2}b_{1}^{2}e_{1}P_{1}r_{1}u_{1}^{2}(Ab_{1}P_{1}u_{1}+(P_{1}-K_{1})w_{1})}{K_{1}(Ab_{1}u_{1}+w_{1})^{3}}.
\]
For an interior equilibrium to be locally asymptotically stable, the determinant
must be positive, i.e.,

\[
P_{1}>\frac{K_{1}w_{1}}{w_{1}+Ab_{1}u_{1}}.
\]
Substituting the two interior equilibria into this inequality, it is easy to see
that only the equilibrium with the higher plant density $(P_{1+},A_{1+})$ satisfies
the above inequality and it is therefore locally stable, while the other equilibrium
is unstable. The position of the two nullclines in Fig \ref{fig:isoclines2d}b confirms
that $(P_{1+},A_{1+})$ is a stable node and $(P_{1-},A_{1-})$ is a saddle point.

\begin{figure}
\begin{centering}
\includegraphics[width=0.5\paperwidth]{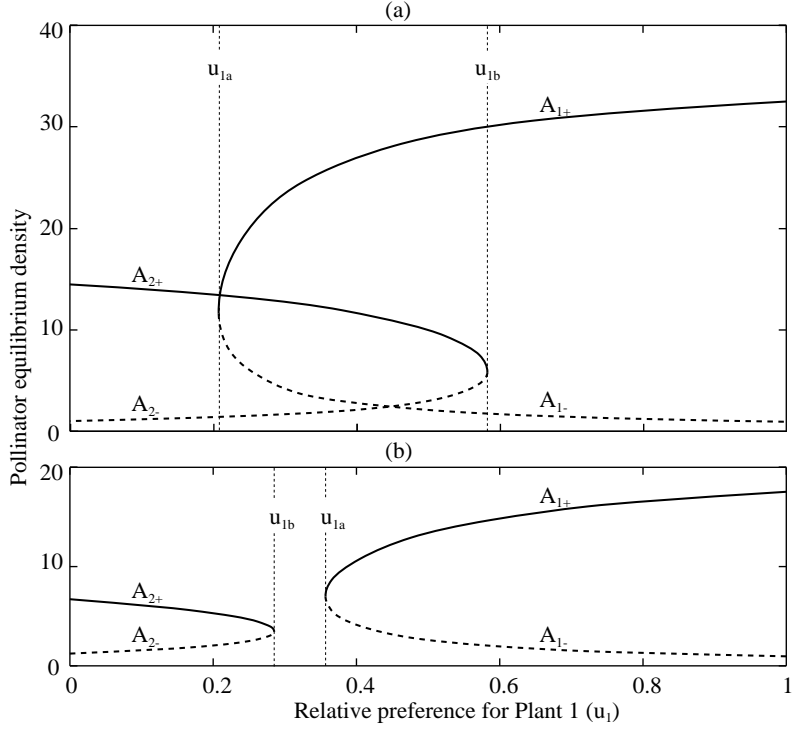} 
\par\end{centering}

\protect\caption{\label{fig:bifPlantAnimal-hilok}Bifurcation plots displaying stable ($A_{i+}$,
solid line) and unstable ($A_{i-}$, dash line) pollinator equilibria under coexistence
with plant $i=1$ or $i=2$ alone. (a) For $K_{1}=K_{2}=60$ inequality (\ref{eq:K1small})
holds. (b) For $K_{1}=K_{2}=35$ inequality (\ref{eq:K1small}) does not hold. The
rest of the parameters are as in Table 1 from the main text.}
\end{figure}

All the results from this section are valid if we ignore plant 1 instead of plant
2, by changing the sub-index 1 to 2. We note that the equilibria $(P_{2\pm},A_{2\pm})$

\noindent 
\[
P_{2\pm}=\frac{b_{2}e_{2}K_{2}(a_{2}r_{2}-m_{2})u_{2}+dr_{2}w_{2}\pm\sqrt{D_{2}}}{2a_{2}b_{2}e_{2}r_{2}u_{2}}
\]
\[
A_{2\pm}=\frac{b_{2}e_{2}K_{2}(a_{2}r_{2}-m_{2})u_{2}-dr_{2}w_{2}\pm\sqrt{D_{2}}}{2b_{2}dr_{2}u_{2}},
\]
and

\begin{equation}
D_{2}=-4b_{2}de_{2}K_{2}m_{2}r_{2}u_{2}w_{2}+(b_{2}e_{2}K_{2}(m_{2}-a_{2}r_{2})u_{2}+dr_{2}w_{2})^{2}.\label{eq:disc2}
\end{equation}

\noindent exist provided

\begin{equation}
u_{1}<1-\frac{dr_{2}w_{2}}{b_{2}e_{2}(\sqrt{a_{2}r_{2}}-\sqrt{m_{2}})^{2}K_{2}}=u_{1b}.\label{eq:u1b_fold}
\end{equation}
Combining (\ref{eq:u1a_fold}) and (\ref{eq:u1b_fold}) we observe that when the
environmental carrying capacity for plant 1 is large so that

\noindent 
\begin{equation}
K_{1}>K_{1}^{*}=\frac{b_{2}de_{2}K_{2}r_{1}(\sqrt{m_{2}}-\sqrt{a_{2}}\sqrt{r_{2}})^{2}w_{1}}{b_{1}e_{1}(\sqrt{m_{1}}-\sqrt{a_{1}}\sqrt{r_{1}})^{2}(b_{2}e_{2}K_{2}(\sqrt{m_{2}}-\sqrt{a_{2}}\sqrt{r_{2}})^{2}-dr_{2}w_{2})}\label{eq:K1small}
\end{equation}

\noindent the two equilibria $(P_{1+},A_{1+})$ and $(P_{2+},A_{2+})$ can coexist
when $u_{1a}<u_{1}<u_{1b}$. If inequality (\ref{eq:K1small}) is reversed, then
$u_{1a}>u_{1b}$ and both single-plant--pollinator equilibria cannot coexist. Fig
\ref{fig:bifPlantAnimal-hilok} shows the dependency of both equilibria, when carrying
capacities are large and small. Numerical bifurcation analysis indicates that equilibria
always come as pairs, an unstable low density equilibrium and a locally stable high
density equilibrium (Fig. \ref{fig:isoclines2d}b).

Because

\noindent 
\[
\frac{\partial}{\partial P_{1}}\left(\frac{1}{P_{1}A}g_{1}(A)\left(1-\frac{P_{1}}{k_{1}(A)}\right)P_{1}\right)+\frac{\partial}{\partial A}\left(\frac{1}{P_{1}A}\left(e_{1}h_{1}(A)P_{1}-d\right)A\right)=
\]
\[
-\frac{(a_{1}b_{1}u_{1}(b_{1}(e_{1}K_{1}+Ar_{1})u_{1}+r_{1}w_{1})}{K_{1}(Ab_{1}u_{1}+w_{1})^{2})}
\]
is negative, the Dulac criterion%
\footnote{J. Hofbauer and K. Sigmund (1998) Evolutionary Games and Population Dynamics, Cambridge
University Press.%
} implies that no limit cycles involving only plant 1 (or plant 2) and the pollinator
exist.

\subsection{Two plant--pollinator coexistence by invasion}

Numerical analysis shows there is also a locally stable interior equilibrium at which
both plants coexist with pollinators. Unfortunately, this equilibrium cannot be expressed
in a closed form and must be analysed numerically. Invasion analysis provides some
partial insight in conditions for species coexistence.

We start with the case where one plant species coexists with pollinators at the interior
locally stable equilibrium and we ask under which conditions the missing plant can
invade. Let us consider the equilibrium $(P_{1+},0,A_{1+})$ at which plant 2 is
missing. This equilibrium exists provided inequality (\ref{eq:u1a_fold}) holds.
In Fig \ref{fig:bifPlantAnimal-hilok} the region of parameters where this equilibrium
exists is to the right of the vertical line at $u_{1a}$. Invasibility of the missing
plant 2 requires

\noindent 
\begin{equation}
g_{2}(A_{1+})\left(1-\dfrac{c_{1}P_{1+}}{k_{2}(A_{1+})}\right)>0,\label{eq:invasionrate}
\end{equation}
i.e., both $g_{2}(A_{1+})$ and $(1-c_{1}P_{1+}/k_{2}(A_{1+}))$ must have the same
sign. Because $g_{2}(A)$ and $k_{2}(A)$ have the same sign for all positive $A$'s
it follows that if $g_{2}$ in (\ref{eq:invasionrate}) is negative, the second term
in parentheses must be positive and (\ref{eq:invasionrate}) cannot hold. Consequently,
the invasion rate can be positive only if $g_{2}(A_{1+})$ is positive, i.e., when
the pollinator abundance at the plant 1--pollinator population equilibrium is high
enough and satisfies

\noindent 
\begin{equation}
A_{1+}>\frac{m_{2}w_{2}}{u_{2}b_{2}(r_{2}a_{2}-m_{2})}\label{eq:inv1}
\end{equation}
to ensure plant 2 positive invasion growth rate. From (\ref{eq:astar}) we can see
that the right-hand-side of this inequality is the threshold pollinator density $A_{2}^{*}$.
In other words, invasion requires that the pollinator density at the equilibrium
$(P_{1+},0,A_{1+})$ must be higher than the minimum mutualistic requirement of the
invader $(A_{2}^{*})$.

Provided (\ref{eq:inv1}) holds, the second term in the right-hand-side of (\ref{eq:invasionrate})
is positive if

\noindent 
\begin{equation}
c_{1}P_{1+}<k_{2}(A_{1+}),\label{eq:2invades}
\end{equation}
i.e., plant 1 equilibrium density cannot be too high to prevent invasion of plant
2, due to strong competition. Substituting the values of $P_{1+}$ and $A_{1+}$
in this inequality, we obtain an inequality in the form $c_{1}<\alpha(u_{1})$, where

\begin{equation}
\alpha(u_{1})=\frac{a_{1}b_{1}r_{1}u_{1}K_{2}\left(2b_{2}u_{2}e_{1}K_{1}m_{1}w_{1}(a_{2}r_{2}-m_{2})-m_{2}w_{2}\left(b_{1}e_{1}K_{1}u_{1}(a_{1}r_{1}-m_{1})-dr_{1}w_{1}-\sqrt{D_{1}}\right)\right)}{a_{2}b_{2}r_{2}u_{2}K_{1}m_{1}w_{1}\left(b_{1}e_{1}K_{1}u_{1}(a_{1}r_{1}-m_{1})+dr_{1}w_{1}+\sqrt{D_{1}}\right)},\label{eq:alpha(u1)}
\end{equation}

\noindent with $D_{1}$ given by (\ref{eq:disc1}).

\noindent Similarly, we obtain invasibility conditions for plant 1 to invade plant
2--pollinator stable interior equilibrium 
\[
A_{2+}>\frac{m_{1}w_{1}}{u_{1}b_{1}(r_{1}a_{1}-m_{1})}
\]
and

\noindent 
\begin{equation}
c_{2}P_{2+}<k_{1}(A_{2+}).\label{eq:1invades}
\end{equation}

\noindent Substituting $P_{2+}$ and $A_{2+}$ in the inequality above, we obtain
an inequality in the form $c_{2}<\beta(u_{1})$, where

\begin{equation}
\beta(u_{1})=\frac{a_{2}b_{2}r_{2}u_{2}K_{1}\left(2b_{1}u_{1}e_{2}K_{2}m_{2}w_{2}(a_{1}r_{1}-m_{1})-m_{1}w_{1}\left(b_{2}e_{2}K_{2}u_{2}(a_{2}r_{2}-m_{2})-dr_{2}w_{2}-\sqrt{D_{2}}\right)\right)}{a_{1}b_{1}r_{1}u_{1}K_{2}m_{2}w_{2}\left(b_{2}e_{2}K_{2}u_{2}(a_{2}r_{2}-m_{2})+dr_{2}w_{2}+\sqrt{D_{2}}\right)}.\label{eq:beta(u1)}
\end{equation}

\noindent In the parameter space showed in the main text, the graph of $\alpha(u_{1})$
is to the right of the $u_{1a}$ vertical line. Below $\alpha$ and right of $u_{1a}$
plant 2 can invade plant 1. And the graph of $\beta(u_{1})$ is to the left of the
$u_{1b}$ vertical line. Below $\beta$ and left of $u_{1b}$ plant 1 can invade
plant 2. Numerical results indicate that when both plants can invade each other,
i.e., when $u_{1a}<u_{1}<u_{1b}$, $c_{1}<\alpha$ and $c_{2}<\beta$, both plants
and the pollinator attain a locally stable equilibrium. In other words we get confirmation
that mutual invasibility implies stable coexistence. However, when mutual invasibility
does not hold, e.g., when only one plant can be a resident, numerical results indicate
more complicated outcomes (see main text).

Because $P_{1+}=k_{1}(A_{1+})$ and $P_{2+}=k_{2}(A_{2+})$, conditions (\ref{eq:2invades})
and (\ref{eq:1invades}) imply that

\begin{equation}
c_{1}c_{2}<Q=\frac{k_{2}(A_{1+})}{k_{1}(A_{1+})}\frac{k_{1}(A_{2+})}{k_{2}(A_{2+})}.\label{eq:c1c2<1}
\end{equation}
This inequality is similar to the competitive exclusion principle%
\footnote{Gause, G. F. (1934) The Struggle for Existence, Williams \& Wilkins.%
} which states that two competing species can coexist only when $c_{1}c_{2}<1,$ i.e.,
when the inter-specific competition is weaker when compared to intra-specific competition.
In the above inequality the right-hand-side $(Q)$ is not equal to 1, but it depends
on the pollinator densities in both single-species--pollinator equilibria. Thus,
this inequality generalises the competitive exclusion principle to a mutualistic--competitive
system with two plants sharing a pollinator. 
\end{document}